\documentclass[aoas,nameyear,MSNbibl,seceqn,dvips]{arximspdf}
\usepackage{graphicx}

\doi{10.1214/11-AOAS455}
\volume{5}
\issue{3}
\pubyear{2011}
\firstpage{1780}
\lastpage{1815}

\makeatletter

\def\RPV{\mathrm{RPG}}
\def\MSPE{\mathrm{MSPE}}
\def\bfb{{\bolds{\beta}}}
\def\bfy{\mathbf{y}}
\def\bmu{{\bolds{\mu}}}
\def\bgamma{{\bolds{\gamma}}}
\def\bfx{\mathbf{x}}
\def\bfz{\mathbf{z}}
\def\DE{\operatorname{DE}}
\def\PVE{\mathrm{PVE}}

\def\bptnote#1{}

\def\@bmisc[#1]{%
  \get@battribute{unstr}%
  \common@pub@types%
  \let\bauthor\bbl@bauthor%
  \let\bhowpublished\@firstofone%
  \def\borganization##1{{\bauthor@style ##1}}%
}

\def\pit{\mathsf{p}}

\makeatother

\begin{document}
\begin{frontmatter}

\title{Bayesian variable selection regression for genome-wide
            association studies and other large-scale problems}
\runtitle{Bayesian variable selection regression for GWAS}

\begin{aug}
\author[A]{\fnms{Yongtao} \snm{Guan}\corref{}\ead[label=e1]{yongtaog@bcm.edu}}
\and
\author[B]{\fnms{Matthew} \snm{Stephens}\ead[label=e2]{mstephens@uchicago.edu}\thanksref{tt1}}
\runauthor{Y. Guan and M. Stephens}
\affiliation{University of Chicago}
\address[A]{
Departments of Pediatrics\\
\quad  and Molecular and Human Genetics\\
Baylor College of Medicine\\
One Baylor Plaza\\
USDA Children's Nutrition Research Center\\
1100 Bates St., Ste. 2070\\
Houston, Texas 77030\\
USA\\
\printead{e1}} 
\address[B]{
Department of Statistics\\
University of Chicago\\
Eckhart Hall Room 126\\
5734 S. University Avenue\\
Chicago, Illinois 60637\\
USA\\
\printead{e2}}
\end{aug}
\thankstext{tt1}{Supported by NIH Grants HG02585 and HL084689.}

\received{\smonth{12} \syear{2009}}
\revised{\smonth{12} \syear{2010}}

%
\begin{abstract}
We consider applying Bayesian Variable Selection Regression, or BVSR,
to genome-wide association studies
and similar large-scale regression problems. Currently, typical
genome-wide association studies measure hundreds of thousands, or
millions, of
genetic variants (SNPs), in thousands or tens of thousands of
individuals, and attempt to identify regions harboring SNPs
that affect some phenotype or outcome of interest. This goal can
naturally be cast as a variable selection regression problem, with the
SNPs as the covariates in the regression. Characteristic features of
genome-wide association studies include the following:
(i) a focus primarily on identifying relevant variables, rather than on
prediction; and
(ii) many relevant covariates may have tiny effects, making it effectively
impossible to confidently identify the complete ``correct'' subset of
variables. Taken together, these factors put a premium on having
interpretable measures of confidence for individual covariates being
included in the model, which
we argue is a strength of BVSR compared with alternatives
such as penalized regression methods. Here we focus primarily on
analysis of quantitative phenotypes, and
on appropriate prior specification for BVSR in this setting,
emphasizing the
idea of considering what the priors imply about the total proportion of
variance in outcome explained by relevant covariates.
We also emphasize the potential for BVSR to estimate this proportion of
variance explained, and
hence shed light on the issue of ``missing heritability'' in genome-wide
association studies.
More generally, we demonstrate that,
despite the apparent computational challenges, BVSR can provide useful
inferences in
these large-scale problems, and in our simulations produces better
power and predictive performance compared with
standard single-SNP analyses and the penalized regression method LASSO.
Methods described here are implemented in a software package,
pi-MASS, available from the Guan Lab website \url{http://bcm.edu/cnrc/mcmcmc/pimass}.
\end{abstract}

%

\end{frontmatter}

\section{Introduction} \label{sec:intro}
The problem of identifying relevant covariates in a regression model,
sometimes known as variable selection,
arises frequently in many fields.
As computational and data-collection technologies have developed, the
number of covariates typically measured in
these kinds of problems has steadily increased, and it is now not
unusual to come across data sets involving
many thousands or millions of covariates. Here we consider one
particular setting where data sets of this size are common:
genome-wide association studies (GWAS).

Current typical GWAS [e.g., \citet{wtccc07}]
measure hundreds of thousands, or millions, of
genetic variants (typically Single Nucleotide Polymorphisms, or SNPs),
in hundreds, thousands, or tens of thousands of individuals, with the
primary goal being to identify which regions of the genome harbor SNPs
that affect some phenotype or outcome of interest. While many GWAS are
case-control studies, here we focus primarily on the computationally-simpler
setting where a continuous phenotype has been measured on
population-based samples, before briefly considering the challenges of
extending these methods to binary outcomes.

Most existing GWAS analyses are ``single-SNP'' analyses, which simply
test each SNP,
one at a time, for association with the phenotype. Strong associations
between a SNP and the phenotype are interpreted as indicating that
SNP, or a nearby correlated SNP, likely affects phenotype.
The primary rationale for GWAS is the idea that, by examining these
SNPs in more detail---for example, examining which genes they are in
or near---we may glean important insights into the biology of the
phenotype under study.

In this paper we examine the potential to apply Bayesian Variable
Selection Regression (BVSR) to GWAS (or other similar
large-scale problems). Variable selection regression provides a very
natural approach to analyzing GWAS: the phenotype is treated as the
regression response, SNPs become regression covariates, and the goal of
identifying genomic regions likely to harbor SNPs affecting phenotype
is accomplished
by examining the genomic locations of SNPs deemed likely to have
nonzero regression coefficients. However, BVSR requires the use of
computationally-intensive Markov chain Monte Carlo (MCMC) algorithms,
and, prior to performing this work,
it was unclear to us whether
such algorithms could produce reliable results in a practical
time-frame for problems as large as a typical GWAS. One important
contribution of this paper is to show that, even using relatively
simple MCMC algorithms, BVSR can indeed produce useful inferences in
problems of this size. Another important contribution is to discuss
\textit{how} BVSR should be used for GWAS analysis, with particular focus
on choice of appropriate prior distribution. Further, and perhaps most
importantly, we
give reasons \textit{why} one might want to use BVSR to analyze GWAS---rather than less computationally-demanding approaches such as
single-SNP analyses, or penalized regression \mbox{approaches} such as LASSO
[\citet{tibshirani96}]---by emphasizing qualitative advantages of BVSR
in this context.
In particular, we emphasize that, unlike penalized regression approaches,
BVSR naturally produces easily-interpretable measures of confidence---specifically, posterior probabilities---that individual covariates
have nonzero regression coefficients.
This is a particularly important advantage in GWAS because the primary
goal of the analysis is to identify such covariates, and to use these
identifications to learn about underlying biology (in contrast to other
settings where prediction may be the primary goal).

Although our work is motivated by GWAS,
many of the ideas and results should be of more general interest. In
brief, the key elements are as follows:
\begin{itemize}
\item We demonstrate that BVSR can be practical for
large problems involving hundreds of thousands of covariates and
thousands of observations.
\item We introduce some new ideas for prior specification in BVSR. In
particular, we emphasize the benefits of focusing on what the priors
imply about the total proportion of variance in response explained by
relevant covariates (henceforth abbreviated as PVE).
We note that standard approaches to prior specification in BVSR, which
put the same priors on the regression coefficients irrespective of how
many covariates are included in the model, imply that models with many
relevant covariates are likely to have much larger PVE than models with
few relevant covariates. We propose a simple alternative prior that
does not make this potentially undesirable assumption, and has the
intuitively appealing property that it applies stronger shrinkage in
more complex models (i.e., models with more relevant covariates).
\item We emphasize the potential for BVSR to estimate the
total amount of signal in a data set, specifically the PVE, even when
there is insufficient information to reliably identify all relevant covariates.
As a result, BVSR has the potential to shed light on the so-called
``missing heritability'' observed in many GWAS [\citet{maher08}; \citet{heightng}].
\item We compare and contrast BVSR with a penalized-regression
approach, the LASSO [\citet{tibshirani96}]. Despite the
considerable literature on both BVSR and penalized regression, there
exist few
comparisons (either qualitative or quantitative) of these two
approaches. We chose the LASSO
as a representative of penalized regression approaches both because of
its popularity and because previous papers have applied
it to the specific context of GWAS [e.g., \citet
{hoggartetal08}; \citet{wuchenetal09}].
In our limited simulation study BVSR outperforms LASSO in terms of
predictive performance.
In addition, we emphasize the qualitative advantage of BVSR over LASSO,
and other penalized regression methods, that it produces posterior
probabilities for each covariate having a nonzero regression
coefficients. This qualitative advantage seems more fundamental, since
predictive performance of different methods may vary depending on the
underlying assumptions.
\end{itemize}

The remainder of the paper is organized as follows. In Section \ref
{sec:modelpriors} we describe BVSR and our choice of priors. In Section
\ref{sec:ci} we discuss computation and inference, including Markov
chain Monte Carlo algorithms used, and a Rao--Blackwellization approach
to estimating the marginal posterior inclusion probability for each
covariate. Section \ref{sec:goals} reviews our main goals in applying
BVSR to GWAS. In Section~\ref{sec:res} we examine, through
simulations, the effectiveness of BVSR for various tasks, including
estimating the PVE, prediction, and identifying relevant covariates.
For some of these tasks we compare BVSR with LASSO and single-SNP
analyses. We also illustrate BVSR on a GWAS for C-reactive protein. In
Section \ref{sec:binary} we briefly consider the challenges of
extending our methods to deal with
binary phenotypes. Finally, in Section~\ref{sec:discussion} we discuss
some limitations and pitfalls of BVSR as we have applied it in this
context, and potential future directions.\looseness=1

\section{Models and priors} \label{sec:modelpriors}

This section introduces notation and specifies the details of
our BVSR model and priors used. Our formulation up~to~Sec\-tion~\ref
{subsec:novel}
is in the same vein as much previous work on BVSR,
but with particular emphasis on putting priors on hyperparameters that
are often considered fixed and known.
Key relevant references include \citet
{mitchellbeauchamp88}, \citet{georgemcculloch93}, \citet{smithkohn96}, \citet{rafteryetal97} and \citet{brownetal02};
see also \citet{miller02} and \citet{ohara09} for more background
and references.

We consider the standard normal linear regression
%
\begin{equation}
\bfy| \bmu, \bfb, X,    \tau \sim N_n(\bmu+ X \bfb, \tau^{-1}I_n),
\end{equation}
relating a response variable $\bfy$ to covariates $X$. Here $\bfy$ is
an $n$-vector of observations on $n$ individuals, $\bmu$ is an
$n$-vector with components all equal to the same scalar $\mu$, $X$ is
an $n$ by $p$ matrix of covariates,
$\bfb$ is a $p$-vector of regression coefficients, $\tau$ denotes the
inverse variance of the residual errors, $N_n(\cdot,\cdot)$ denotes
the $n$-dimensional multivariate normal distribution and $I_n$ the $n$
by $n$ identity matrix. The variables $\bfy$
and $X$ are observed, whereas $\bmu, \bfb$, and $\tau$ are
parameters to be inferred.
In more detail, $\bfy= (y_1,\ldots ,y_n)$, where $y_i$ is the measured response
on individual $i$, and $X=(\bfx_{\cdot1},\ldots ,\bfx_{\cdot p})$,
where $\bfx_{\cdot j}=(x_{1j},\ldots ,x_{nj})^T$ is a column vector
containing the observed values of the $j$th covariate. For example, in
the context of a GWAS, $y_i$ is the measured phenotype of interest in
individual $i$, and $x_{ij}$ is the genotype of individual $i$ at SNP
$j$, typically coded as 0, 1 or 2 copies of a particular reference
allele. [By coding the genotypes as 0, 1, or 2, we are assuming an
additive genetic model. It would be straightforward to include dominant
and recessive effects by adding another covariate for each SNP, as in
\citet{servinstephens07}, e.g., although this would increase
computational cost.]

In many contexts, including GWAS, the number of covariates is very
large---and, in particular, $p \gg n$---but only a small subset of the covariates are expected to be associated
with the response (i.e., have nonzero $\beta_j$). Indeed, the main
goal of GWAS is to identify these relevant covariates.
To this end, we define a vector of binary indicators $\bgamma=(\gamma
_1,\ldots ,\gamma_p) \in\{0,1\}^p$
that indicate which elements of $\beta$ are nonzero. Thus,
%
\begin{equation} \label{model}
\bfy| \bgamma, \mu, \tau, \bfb,    X \sim N_n(\bmu+ X_\bgamma\bfb
_\bgamma, \tau^{-1} I_n),
\end{equation}
where $X_\bgamma$ denotes the design matrix $X$ restricted to those
columns $j$ for which $\gamma_j=1$, and $\bfb_\bgamma$ denotes a
corresponding vector
of regression coefficients.
In general, for observational studies one would be reluctant to
conclude any causal
interpretation for $\bgamma$, but in the context of GWAS,
it is usually reasonable to interpret $\gamma_j=1$ as indicating that
SNP $j$,
or an unmeasured SNP correlated with SNP $j$, has a causal (functional)
affect on $\bfy$.
This is because in GWAS reverse causation is generally implausible
(phenotypes cannot causally affect genotype, since genotype comes first
temporally), and there
are few potential unmeasured confounders that could affect both
genotype and $\bfy$ [\citet{smithebrahim03}]. A~well-documented
exception to this is population structure; here we assume that this has
been corrected for prior to analysis, for example, by letting $\bfy$
be the residuals from regressing the observed phenotype values against
measures of population structure, obtained, for example, by model-based
clustering [\citet{pritchardetal00}] or principal components analysis
[\citet{priceetal06}].

Taking a Bayesian approach to inference, we put priors on the parameters:
%
\begin{eqnarray} \label{prior:mutau}
\tau& \sim&\operatorname{Gamma}(\lambda/2,\kappa/2), \\
\mu| \tau& \sim& N(0,\sigma_\mu^2/\tau), \\
\gamma_j & \sim&\operatorname{Bernoulli}(\pi), \\\label{eqn:betaprior}
\bfb_\bgamma| \tau, \bgamma
& \sim& N_{|\bgamma|} \bigl(0, (\sigma_a^2/\tau) I_{|\bgamma|}
\bigr),  \\
\bfb_{-\bgamma} | \bgamma& \sim&\delta_0,
\end{eqnarray}
where $|\bgamma| := \sum_j \gamma_j$, $\bfb_{-\bgamma}$ denotes
the vector of $\bfb$ coefficients for which $\gamma_j=0$, and $\delta
_0$ denotes a point mass on 0.
Here $\pi, \sigma_a, \lambda, \kappa$, and $\sigma_\mu$ are
hyperparameters. The hyperparameters $\pi$ and $\sigma_a$ have
important roles, with $\pi$ reflecting the sparsity of the model, and
$\sigma_a$ reflecting the typical size of the nonzero regression
coefficients. Rather than setting these hyperparameters to prespecified
values, we place priors on them, hence allowing their values to be
informed by the data; the priors used are detailed below. (Later we
will argue that this ability to infer $\pi$ and $\sigma_a$ from the
data is an important advantage of analyzing all SNPs simultaneously,
rather than one at a time.) The remaining hyperparameters are less
critical, and, in practice, we consider the posterior distributions for
which $\sigma_\mu^2 \rightarrow\infty$ and $\nu,\kappa\rightarrow
0$, which has
the attractive property that the resulting relative marginal
likelihoods for~$\bgamma$ are invariant to shifting or scaling of
$\bfy$. Thus, for example, inference of which genetic variants are
associated with height would be unaffected
by whether height is measured in meters or inches. [Taking these limits
is effectively equivalent to using the improper prior $p(\mu,\tau)
\propto1/\tau$, but we prefer to formulate proper priors and take
limits in their posteriors, to verify sensible limiting behavior.]

The parameter $\pi$ controls the sparsity of the model, and where the
appropriate level of
sparsity is uncertain  a priori, as is typically the case, it
seems important to specify
a prior for $\pi$ rather than fixing it to an arbitrary value. In
GWAS, and probably in many other
settings with extreme sparsity, uncertainty in $\pi$ may span orders
of magnitude: for example, there could be just a few relevant
covariates or hundreds. In this case a uniform prior on~$\pi$ seems
inappropriate, since this
would inevitably place most of the prior mass on larger numbers of
covariates (e.g., uniform on $10^{-5}$ to $10^{-3}$ puts about $90\%$
probability on $>$10$^{-4}$). Instead, we put a uniform prior on $\log
\pi$:
%
\begin{equation}\label{prior:s}
\log{\pi} \sim U(a, b),
\end{equation}
where $a = \log(1/p)$ and $b = \log(M/p)$, so the lower and upper
limits on~$\pi$ correspond, respectively, to an expectation of $1$ and
$M$ covariates in the model.
In applications here we used $M=400$, with this arbitrary limit being
imposed partly due to computational considerations (larger $M$ can
increase computing time considerably). The assumption of a uniform
distribution is, of course, somewhat artificial but has the merit of
being easily interpretable. An alternative, which may be preferable in
some settings, would be a normal prior on $\log({\pi}/{(1-\pi)})$.

The above formulation, with the exception of our slightly nonstandard
prior on $\pi$, follows previous work. However, since many
formulations of BVSR differ slightly from one another, we
now comment on some of the choices we made:
\begin{longlist}[(5)]
\item[(1)] We chose, in (\ref{eqn:betaprior}), to put independent priors on
the elements of $\bfb_\bgamma$. An alternative common choice is
Zellner's $g$-prior [\citet{zellner86}; \citet{zellner88}], which assumes
correlations among the regression coefficients mimicking the
correlations among covariates,
\[
\bfb_\bgamma\sim N_{|\bgamma|} \biggl(0, \frac{g}{\tau} X^t_\bgamma
X_\bgamma \biggr).
\]
For GWAS we prefer the independent priors because
we view the $\bfb$'s as reflecting causal effects of $X$ on $\bfy$,
and there seems to be no
good reason to believe that the correlation structure of causal effects
will follow that of the SNPs.
\item[(2)] Some authors center each of the vectors $\bfy$ and $\bfx_{\cdot
1},\ldots ,\bfx_{\cdot p}$ to have mean $0$, and set $\mu=0$. This
approach yields the same posterior on $\bgamma$ as our limiting prior
on $\mu$ (derivation omitted), and simplifies calculations,
and so we use it henceforth.
\item[(3)] It is common in variable selection problems to scale the
covariates $\bfx_{\cdot1},\ldots ,\bfx_{\cdot p}$ to each have unit
variance, to avoid problems due to different variables being measured
on different scales. In GWAS these covariates are measured on the same
scale, being counts of the reference allele,
and so we do not scale the covariates in this way in our examples.
However, one could so scale them,
which would correspond to a prior assumption that SNPs with less
variable genotypes (i.e., those with a lower minor allele frequency)
have larger effect sizes; see \citet{wakefield08} for relevant discussion.
\item[(4)] The priors assume that the $\beta_j$ are exchangeable, and, in
particular,
that all covariates are, a priori, equally plausible candidates to
affect outcome $\bfy$. In the context of a GWAS, this assumption means
we are ignoring information that might make some SNPs better candidates
for affecting outcome than others. Our priors also ignore the fact that
functional SNPs may tend to cluster near one another in the genome.
These choices were made purely for simplicity; one attractive feature
of BVSR is that one could modify the priors to incorporate these types
of information, but we leave this to future work.
\item[(5)] Some formulations of BVSR use a similar ``sparse'' prior, where
the marginal prior on $\beta_j$ is a mixture of a point mass at 0 and
a normal distribution, whereas others [e.g., \citet
{georgemcculloch93}] instead use a~mixture of two normal
distributions, one with a substantially larger variance than the other.
The sparse formulation seems computationally advantageous in large
problems because sparsity facilitates certain operations (e.g.,
integrating out $\bfb$ given $\bgamma$).
\end{longlist}

\subsection{\texorpdfstring{Novel prior on $\sigma^2_a$}{Novel prior on sigma^2_a}} \label{subsec:novel}

While the above formulation is essentially standard and widely used,
there is considerable variability in how different authors treat the
hyperparameter $\sigma_a$. Some fix it to an essentially arbitrary
value, while others put a prior on this parameter. Several different
priors have been suggested, and the lack of consensus among different
authors may reflect the fact that
most of them seem not to have been given a compelling motivation or
interpretation.
Here we suggest a way of thinking about this prior that we believe aids
interpretation, and hence appropriate prior specification.
Specifically, we suggest focusing on what the prior implies about the
proportion of variance in $\bfy$ explained by $X_{\bgamma}$ (the PVE).
For example, almost all priors we have seen previously in this context
assume independence of $\pi$ and $\sigma_a$, which implies
independence of $\bgamma$ and $\sigma_a$. While this assumption may
seem natural initially, it implies that more complex models are
expected to have substantially higher PVE. In our application this
assumption does not capture our prior beliefs. For example, it seems
quite plausible a priori that there could be either a large number of
relevant covariates
with small PVE, or a small number of covariates with large PVE.

Here we suggest specifying a prior on $\sigma^2_a$ given $\bgamma$ by
considering the induced prior on the PVE, and, in
particular, by making this induced prior relatively flat in the range
of $(0,1)$.
To formalize this,
let $V(\bfb,\tau)$ denote the empirical variance of $X\bfb$ relative
to the residual variance $\tau^{-1}$:
%
\begin{equation}
V(\bfb, \tau) := \frac{1}{n}\sum_{i=1}^n {[{(X \bfb)}_i]^2} \tau,
\end{equation}
where this expression for the variance assumes that the covariates have
been centered, and so $X\bfb$ has mean 0.
Then the
total proportion of variance in $\bfy$ explained by $X$ if the true
values of the regression coefficients are $\bfb$ is given by
%
\begin{equation}\label{eqn:H}
\PVE(\bfb, \tau) := V(\bfb,\tau) / \bigl(1+ V(\bfb,\tau)\bigr).
\end{equation}
Our aim is to choose a prior on $\bfb$ given $\tau$ so that the
induced prior on $\PVE(\bfb, \tau)$ is approximately uniform. To do
this, we exploit the fact that the expected value of $V(\bfb,\tau)$
(with expectation being taken over $\bfb| \tau$) depends in a simple
way on $\sigma_a$:
%
\begin{equation} \label{eqn:v}
v(\bgamma,\sigma_a) := E[V(\bfb,\tau) | \bgamma, \sigma_a, \tau]
= \sigma_a^2 \sum_{j\dvtx \gamma_j=1} {s_j },
\end{equation}
where $s_j = \frac{1}{n}\sum_{i=1}^n{x_{ij}^2}$ is the variance of
covariate $j$. Define
%
\begin{equation} \label{eqn:h}
h(\bgamma, \sigma_a) = v(\bgamma, \sigma_a) / \bigl(1+v(\bgamma, \sigma_a)\bigr).
\end{equation}
Intuitively, $h$ gives a rough guide to the expectation of PVE for a
given value of $\bgamma$ and $\sigma_a$.
(It is not precisely the expectation since it is the ratio of
expectations, rather than the expectation of the ratio.)
To accomplish our goal of putting approximately uniform prior on $\PVE
$, we specify a uniform prior on $h$, independent of $\bgamma$, which
induces a prior on $\sigma_a$ given $\bgamma$ via the relationship
%
\begin{equation}\label{eqn:sigma}
\sigma_a^2(h,\gamma) = \frac{h}{1-h}\frac{1}{\sum_{j\dvtx \gamma_j=1}{s_j}}.
\end{equation}
In all our MCMC computations, we parameterize our model in terms of
$(h, \bgamma)$, rather than $(\sigma_a, \bgamma)$. Note that the
induced prior on $\sigma_a^2$ is diffuse:
if $Z= h/ (1-h)$, and $h \sim U(0,1)$, then $Z$ has a probability
density function $f(z) = 1/(1+z)^2$, which is heavy tailed.

Our prior on $\sigma_a$ has interesting connections with the prior
suggested by \citet{liangetal08}. While
\citet{liangetal08} use a $g$ prior, if we consider the case where
the covariates are orthogonal with variances $s_j= 1$, then their
parameter $g$ is
effectively equivalent to our $n \sigma_a^2$. They suggest putting a~$\operatorname{Beta}(1, a/2 -1)$ prior on $g/(1+g)$, with $a = 3$ or $4$;
the case $a=4$ is uniform on $g/(1+g)$, or in our notation uniform on
$n \sigma_a^2 / (1+n\sigma_a^2)$. In contrast, our prior is uniform
on $|\bgamma| \sigma_a^2/ (1+ |\bgamma| \sigma_a^2)$. Thus, our
$\sigma_a$ is effectively $n/|\bgamma|$ times the value of $\sigma
_a$ from \citet{liangetal08}, and so our $\sigma_a$ is larger than
theirs (implying less shrinkage), provided that the number of relevant
covariates~$|\bgamma|$ is less than $n$. Qualitatively, perhaps the
main difference between the priors is that our prior applies less
shrinkage (larger $\sigma_a$) in
simpler models, which seems intuitively appealing.

Of course, choice of appropriate prior distributions may vary according
to context, and we do not argue that the
prior used here is universally superior to other choices. However, we
do believe the priors
outlined above are suitable for general use in most GWAS applications.
In addition, we emphasize
that these priors incorporate two principles that we believe should be
helpful more generally:
first, it seems preferable to place prior distributions on the
hyperparameters $\pi$ and $\sigma_a$, rather than fixing
them to specific values, as this provides the potential to learn about
them from the data; second, when
comparing priors for $\sigma_a$, it is helpful to consider what the
priors imply about $\PVE$.

\section{Computation and inference} \label{sec:ci}

We use Markov chain Monte Carlo to obtain samples from the posterior
distribution of $(h, \pi, \bgamma)$ on the product space $(0,1)
\times(0, 1) \times\{0,1\}^p$, which is given by
%
\begin{equation} \label{post:hsg}
\pit(h, \pi, \bgamma| \bfy) \propto\pit(\bfy| h, \bgamma) \pit
(h) \pit(\bgamma| \pi) \pit(\pi).
\end{equation}
Here we are exploiting the fact that the parameters $\bfb$ and $\tau$
can be integrated out analytically to
compute the marginal likelihood $\pit(\bfy| h, \bgamma)$. For each
sampled value of $h,\bgamma$ from this posterior, we also obtain
samples from the posterior distributions of $\bfb$ and $\tau$ by
sampling from their conditional distributions given $\bfy,\bgamma,h$.

Our Markov chain Monte Carlo algorithm for sampling $h,\pi,\bgamma$
is detailed in Appendix~\ref{appmA}. In brief, it is a Metropolis--Hastings
algorithm [\citet{metropolisetal53}; \citet{hastings70}], using a simple local
proposal to jointly update $h, \pi,\bgamma$.
In particular, it explores the space of covariates included in the
model, $\bgamma$, by proposing to add, remove, or switch
single covariates in and out of the model. To improve computational
performance, we use three strategies. First, in addition to the local
proposal moves, we sometimes make a longer-range proposal by
compounding randomly many local moves. This technique, named
``small-world proposal,'' improves the theoretical convergence rate of
the MCMC scheme [\citet{guankrone07}].
Second, and perhaps more importantly, when proposing new values for
$\bgamma$, and specifically when
proposing to add a variable to the model, we
focus more attention on those
covariates with the strongest \textit{marginal}
associations. This idea is related to the sure independence screen
[\citet{fanlv08}] (SIS), which
uses marginal associations as an initial screening step. However,
it is a ``softer'' use of these marginal associations than the SIS,
because every variable continues to have
positive probability of being proposed. Simulations (not shown) show
that taking account
of the marginal associations in this way dramatically increases the
acceptance rate compared to a proposal that ignores the marginal
associations. Finally, when estimating quantities of interest, we make
use where possible of Rao--Blackwellization techniques [\citet
{casellarobert96}], detailed below, to reduce Monte Carlo variance.\looseness=1

We note that our computational scheme is relatively simple, and one can
create data sets where it will perform poorly, for example,
multiple correlated covariates that are far apart along a chromosome,
where an efficient algorithm would require careful joint updates of the
$\gamma_i$ values for those correlated covariates. (In our current
implementation, swap proposals only apply to SNPs that are close to one
another in the genome, which is motivated by the fact that correlations
decay quickly with respect to distance between SNPs.)
However, our main focus in this paper is not on producing a
computational scheme that will deal with difficult situations
that might arise, but rather on prior specification, and to provide an
initial assessment of the potential for BVSR to be applied to
large-scale problems. Indeed, we hope
that our results stimulate more research on the challenging
computational problems that can occur in applying BVSR to GWAS
and similar settings.\looseness=1


\subsection{Posterior inclusion probabilities via Rao--Blackwellization}

In the context of GWAS, a key inferential question is which covariates
have a high
probability of being included in the model. That is, we wish to compute
the \textit{posterior inclusion probability} (PIP) of the $j$th
covariate, $\Pr(\gamma_j=1 | \bfy)$. Although
one could obtain a simple Monte Carlo estimate of this probability by
simply counting the proportion of MCMC samples for which
$\gamma_j=1$, this estimator may have high sampling variance. To
improve precision, we instead use the
Rao--Blackwellized estimate,
%
\begin{equation}\label{post:inclusion}
\Pr(\gamma_j=1 | \bfy) \approx(1/M) \sum_{i =1}^M \Pr\bigl(\gamma_j=1
| \bfy,
\bgamma^{(i)}_{-j}, \bfb^{(i)}_{-j}, \tau^{(i)}, h^{(i)}, \pi^{(i)}\bigr),
\end{equation}
where $\bgamma^{(i)},\bfb^{(i)},\tau^{(i)},h^{(i)},\pi^{(i)}$
denote the $i$th MCMC sample from the posterior distribution of these
parameters given $\bfy$, and $\bgamma_{-j}$ and $\bfb_{-j}$ denote
the vectors $\bgamma$ and $\bfb$ excluding the $j$th coordinate.
The probabilities that are being averaged here essentially involve
simple univariate regressions of residuals against
covariate $j$, and so are fast to compute for all $j$ even when~$p$ is
very large. Details are given in Appendix~\ref{appmB}.

\subsection{Estimating proportion of variance explained}

To perform inference on the total proportion of variance in $\bfy$
explained by measured covariates, we use samples from the posterior
distribution of $\PVE(\bfb,\tau)$, which is defined\vspace*{1pt} at equation
(\ref{eqn:H}). These posterior samples are obtained by simply
computing $\PVE(\bfb^{(i)},\tau^{(i)})$ for each sampled value of
$\bfb,\tau$ from our MCMC scheme.

\subsection{Predicting future exchangeable observations}

Given observed covariates $x_{n+1}$ for a future individual, we can predict
a value of $y_{n+1}$ for that individual by
%
\begin{equation}
E(y_{n+1} | \bfy) = x_{n+1} E(\bfb| \bfy).
\end{equation}
To estimate $E(\bfb| \bfy)$, we use the Rao--Blackwellized estimates
%
\begin{equation} \label{eqn:betarb}
E(\beta_j | \bfy) \approx(1/M) \sum_{i=1}^M E\bigl(\beta_j | \gamma
_j=1, \bfy, \theta^{(i)}_{-j}\bigr) \Pr\bigl(\gamma_j=1 | \bfy, \theta^{(i)}_{-j}\bigr).
\end{equation}
Expressions for the two terms in this sum are given in Appendix~\ref{appmB}.

\subsection{Assessing predictive performance} \label{subsec:prediction}

Suppose that we estimate $\bfb$ to be
$\hat{\bfb}$. One way to assess the overall quality of this estimate
is to ask how well
it would predict future observations, on average. Motivated by this, we
define the mean squared prediction error ($\MSPE$):
%
\begin{equation} \label{eqn:loss}
\MSPE(\hat{\bfb}; \bfb, \tau)
= E(X \hat{\bfb} - \bfy)^2
= \sum_{j=1}^m {s_j (\hat{\bfb} - \bfb)^2 + 1/\tau},
\end{equation}
where $\bfb$ is the true value of the parameter, and $s_j$ is the
variance of $\bfx_{\cdot j}$, defined at (\ref{eqn:v}).

The $\MSPE$ has the disadvantage that its scale depends on the units
of measurement of $\bfy$. Hence, we define a \textit{relative
prediction gain}, $\RPV$, which contrasts the $\MSPE$
from an estimated $\bfb$ with the
prediction loss from simply predicting the mean of $\bfy$ for each
future observation ($\MSPE_0$) and to the prediction error attained by
the true value of $\bfb$ ($\MSPE_{\mathrm{opt}} = 1/\tau$):
%
\begin{equation} \label{eqn:rpv}
\RPV= \frac{\MSPE_0 - \MSPE(\hat{\bfb})}{\MSPE_0 - \MSPE_{\mathrm{opt}}}.
\end{equation}
The $\RPV$ does not depend on $\tau$ or on the scale of measurement
of $\bfy$, and indicates what proportion
of the extractable signal we are successfully obtaining from the data.
For example, if the total proportion of variance in $\bfy$ explained
by $X_\bgamma$
is 0.2, then an $\RPV$ of 0.75 indicates that we are effectively able to
extract three-quarters of this signal, leaving approximately 0.05 of
the variance
in $\bfy$ ``unexplained.''
Note that $\RPV=0$ if the prediction performs as well as the mean,
and $\RPV=1$ if the prediction performs as well as the true value of
$\bfb$.
If $\RPV<0$, then the prediction
is worse than simply using the mean, effectively indicating a problem
with ``overfitting.''

\section{Goals and expectations}\label{sec:goals}

At this point it seems helpful to review what we are attempting to achieve,
and why it might be achievable despite the apparent severity of the
computational burden.
In brief, our primary goal is to extract more information from signals
that exist in these data, particularly marginal associations, than do
standard single-SNP analyses that test each SNP, one at a time, for
association with phenotype. (The vast majority of GWAS published so far
restrict themselves to such single-SNP analyses.) One of the main
difficulties in single-SNP analysis is to decide how confident one
should be that individual SNPs are truly associated with the phenotype.
This difficulty stems from the fact that confidence should depend on
the unknown values of $\pi$ and $\sigma_a$.
In a single-SNP analysis one must make assumptions, either implicitly
or explicitly, about these parameters.
An important aim of our approach is to instead estimate these
parameters from the data, and hence provide more data-driven
estimates of confidence for each SNP being associated with phenotype.
To get intuition into why the data are informative about $\pi$ and
$\sigma_a$, consider the following examples.
First, suppose that in a GWAS involving 300,000 SNPs,
there are 10 SNPs (in different genomic regions) that show very strong
marginal associations. Then, effectively,
we immediately learn that $\pi$ is likely at least
of the order of $10/300\mbox{,}000$ (and, of course, it may be considerably
higher). Further, the estimated size of the effects at these 10 SNPs
also immediately gives some idea of plausible values for $\sigma_a$
(or, more precisely, for $\sigma_a/\tau$).
Conversely, suppose that in a different GWAS none
of the 300,000 SNPs show even modest marginal associations. This
immediately suggests that either $\pi$ or $\sigma_a/\tau$ (or both)
must be ``small'' (because if both were large, then there would be many
strong effects, and we would have seen some of them).
More generally, we note that the strength of the effect size at the
most strongly associated SNPs immediately puts an upper bound
on what kinds of effect size are possible, and hence an upper bound on
plausible values for $\sigma_a/\tau$.
In essence,
BVSR provides a model-based approach to quantifying these qualitative
ideas, taking account of relevant factors (e.g., sample size)
that affect the amount of information in the data.

Another limitation of single-SNP analyses, at least as conventionally
applied, is that once some
SNPs are confidently identified to be associated with outcome, they are
not controlled for, as they should be, in analysis of subsequent SNPs.
Controlling for SNPs that truly affect phenotype should help in
identifying further such SNPs, and so a second
key aim of our approach is to accomplish this. To see why our approach
should attain this goal,
note that our Rao--Blackwellization procedure for estimating marginal
posterior inclusion probabilities is effectively
a conventional single-SNP analysis that controls for the SNPs currently
in the model. Thus, for example, if we start our MCMC algorithm
from a point that includes the strongest marginal associations, it
effectively immediately accomplishes this second goal.

We note two things we are \textit{not} attempting to do. First, we are not
attempting to identify a
single best model (i.e., combination of SNPs), or to estimate posterior
probabilities for specific models. In this context---and, we would
argue, many other contexts where
BVSR may be appropriate---these goals are of no interest, because the
combination of small effect sizes and $p \gg n$ mean that the posterior
probability on any particular model is going to be very small, and the
chance of identifying the
``correct'' model is effectively zero. Neither are we attempting to
identify combinations of SNPs that interact in a nonadditive
way to affect the phenotype---SNPs that have little marginal signal,
but whose effect is only revealed when they are considered
together in combination with others. While such combinations of SNPs
may exist, and identifying them would be of considerable interest, this
seems considerably more challenging, both statistically and
computationally, than our more modest goals here.

Finally, we note a particular feature of GWAS studies that may make it
easier to obtain useful results from BVSR than in other contexts.
Specifically, correlations among SNPs tend to be highly ``local'': each
SNP is typically correlated with only
a relatively small number of other SNPs that are near to it (linearly
along the DNA sequence), and any two randomly chosen
SNPs are typically uncorrelated with one another. Put another way, the
matrix $X'X$ tends to have a highly banded structure, with large
values clustering near the diagonal. To understand why this is helpful,
note that one of the main potential pitfalls in applying
MCMC to BVSR is that the MCMC scheme may get stuck in a ``local mode''
where a particular covariate
($A$,~say) is included in the model, whereas in fact a different
correlated covariate ($B,$ say) should have been included.
To help avoid getting stuck like this, the MCMC scheme could include
specific steps that propose to interchange correlated covariates (e.g.,
remove $A$ from the model
and add $B$ to the model), and the local correlation structure among
SNPs in a GWAS means that this is easily implemented by simply proposing
to interchange nearby SNPs. Furthermore, and perhaps more importantly,
the local correlation structure means
that getting stuck in such local modes may not matter very much,
because if $A$ and $B$ are correlated,
then they are also almost certainly close to one another in the genome,
and hence implicate a similar set of genes,
and correctly identifying a set of implicated genes is the ultimate
goal of most GWAS analyses.

\section{Simulations and comparisons with other methods} \label{sec:res}

We now present a~variety of simulation results to illustrate features
of our method,
and assess its performance. Because our priors and methodology were
primarily motivated by
GWAS, these simulations are designed to mimic certain features of a
typical GWAS.
These include particularly that $p \gg n$ (in our simulations $p
\approx10\mbox{,}000\mbox{--}300\mbox{,}000$ and $n \approx1\mbox{,}000$), extreme sparsity (in
most of our simulations ${\sim}30$ covariates affect response), and
small effect sizes (most relevant covariates individually explain $< $1\%
 of the variance of $\bfy$).

\subsection{Simulation details} \label{sec:sim}

We performed simulations based on three different genotype data,
including both simulated and real genotypes. The first is simulated
$10\mbox{,}000$ independent SNPs (henceforth 10K), the
second is real genotypes at ${\sim}317\mbox{,}000$ SNPs (henceforth 317K), and
the third is real genotypes at ${\sim}550\mbox{,}000$ SNPs (henceforth 550K).
Both 317K and 550K genotypes closely mimic real GWAS, and comparison
between them can illustrate the scalability of our method. The 10K data
set is helpful for several reasons: smaller simulations run faster;
they allow us to assess methods in a simpler setting
where computational problems are less of an issue; and the independence
of the covariates avoids problems with deciding what is meant by
a~``true association'' when covariates are correlated with one another.\looseness=1

For the 10K data, we simulated genotypes as follows. At each SNP
$j=1,\ldots ,10\mbox{,}000$ the minor allele frequency $f_j$ is drawn from a
uniform distribution on $[0.05, 0.5]$, and then genotypes $x_{ij}$ ($i
= 1,\ldots ,n$) are drawn independently from a $\operatorname{Binomial}(2, f_j)$
distribution. We use $n=1\mbox{,}000$ and $6\mbox{,}000$.

Both 317K and 550K data sets come from an association study performed
by the Pharmacogenomics and Risk of Cardiovascular Disease (PARC)
consortium [\citet{reineretal08}; \citet{barberetal09}]. The 317K genotypes
come from the Illumina 317K BeadChip SNP arrays for 980 individuals and
the 550K genotypes come from the Illumina 610K SNP chip plus a custom
13,680 SNP Illumina \mbox{i-Select} chip in 988 individuals (550K SNPs remain
after QC). We replaced missing genotypes with their posterior mean
given the observed genotypes, which we computed under a Hidden Markov
Model [\citet{scheetstephens06}] implemented in the software package
BIMBAM [\citet{guanstephens08}].

For both the simulated and real genotypes we simulated sets of
phenotypes in the following way. First, we specified a value
of $\PVE$, the total proportion of variance in $\bfy$ explained by
the relevant SNPs, that we wanted to achieve in the simulated data.
Then we randomly selected a set of $30$ ``causal'' SNPs, $C$, and
simulated effect sizes $\beta_j$ for each of these SNPs independently
from an effect size distribution $\mathcal{E}(\cdot)$ (discussed
below). Next we computed the value of $\tau$ that gives the desired
value for $\PVE(\bfb, \tau)$ in equation (\ref{eqn:H}). Finally, we
simulated phenotypes for each individual using $y_i = \sum_{j\in
C}{\beta_j x_{ij}} + N(0,\tau^{-1})$.\looseness=1

Unless otherwise stated, for the 10K SNP data sets we run BVSR for 1
million iterations, and for the 317K and 550K SNP data sets we use 2
million iterations. Run times for each data set varied from a few
minutes to about one day on a single MAC Pro with 3 GHz processor. (Note
that the running time per iteration depends primarily on the inferred
values for $|\bgamma|$, not
the total number of SNPs.)

\subsection{Other methods}\label{subsec:other}

In results presented below we compare our method with two other
methods: simple single-SNP analysis that tests
each SNP one at a time for association with phenotype, and the
penalized regression method LASSO [\citet{tibshirani96}].

For the single SNP analyses we ranked SNPs by their single-SNP Bayes
factors, computed using equation (\ref{eqn:bf}), with $\bgamma$ in
the numerator being the vector with $j$th component 1 and all other
components 0, and averaging over $\sigma_a = 0.4, 0.2$, and 0.1 as in
\citet{servinstephens07}. (Using standard single-SNP $p$ values
instead of Bayes Factors gives very similar performance in terms of
ranking SNPs.)

The LASSO procedure [\citet{tibshirani96}] estimates $\bfb$ by
minimizing the penalized residual sum of squares:
%
\begin{equation}\label{eqn:lasso}
\operatorname{argmin}\limits_{\bfb}  (\bfy- X\bfb)^t (\bfy-X\bfb) + \lambda
\sum_j |\beta_j|.
\end{equation}
For sufficiently large penalties, $\lambda$, LASSO produces sparse
estimates $\hat{\bfb}$. Its main practical advantage over BVSR
appears to be computational: for example, one can efficiently find the
global optimal solution path for $\bfb$ as $\lambda$ varies. To apply
the LASSO procedure, we used the \texttt{lars} package (v. 0.9-7) in R
[\citet{efronetal04}].

\subsection{Inference of PVE, and its relationship to heritability}

The total proportion of variance in $\bfy$ explained by the relevant
covariates $X_\bgamma$, or PVE, is commonly used to summarize the
results of a linear regression. In GWAS the $\PVE$ is, conceptually,
closely related to
the ``heritability'' of the trait, which is widely used, for better or
worse, as a summary of how ``genetic'' the phenotype is. The key
difference between the $\PVE$ and heritability is that the $\PVE$
reflects the optimal predictive accuracy that could be achieved for a
linear combination of the \textit{measured} genetic variants, whereas
heritability reflects the accuracy that could be achieved by \textit{all}
genetic variants.
In recent GWAS, it has been generally observed, across a range of
different diseases and clinical traits, that the proportion of
phenotypic variance explained by ``significant'' genetic variants is
much lower than previous estimates of heritability from family-based
studies [\citet{maher08}]. There are several possible explanations for
this ``missing heritability'': for example, it may be that previous
estimates of heritability are inflated for some reason. However, two
explanations have received
particular attention: some of the missing heritability could reflect
genetic variants that were measured but
simply did not reach stringent levels of ``significance'' in standard analyses,
while other parts of the missing heritability could reflect genetic
variants that were
unmeasured (and not strongly correlated with measured variants).
Because the measured genetic variants
in current GWAS studies are
predominantly ``common'' genetic variants (those with a population
frequency exceeding a~few percent),
the relative contribution of these
two factors is connected to the contentious topic of the relative
contributions of common vs rare variants
to phenotypic variation and disease risk [\citet{pritchard01}].
Comparing the $\PVE$ with heritability should provide some insights
into the relative contributions of these
two factors. For example, at the simplest level, if the $\PVE$ is
almost as big as the heritability, then this suggests that
most phenotypic variation is due to variation at SNPs that are highly
correlated with measured genetic variants,
and perhaps that rare genetic variants, which are usually not strongly
correlated with measured common variants, contribute little
to phenotypic variation.

An important feature of BVSR that allows it to estimate the $\PVE$,
together with measures of confidence, is
its use of \textit{Bayesian model averaging} (BMA) to average over
uncertainty in which covariates are relevant.
This is very different from single SNP analyses and standard penalized
regression approaches,
which typically result in identification of a single set of
potentially-relevant covariates, and
so do not naturally provide estimates of the $\PVE$ that take account
of the fact that this set may
be missing some relevant covariates and include some irrelevant covariates.
Since, as far as we are aware, the ability of BVSR to estimate $\PVE$
has not been examined previously, we performed simulation studies to
assess its potential.

For both real and simulated genotype data (described above), we
simulated 50 independent sets of phenotype data, each containing 30
randomly-chosen ``causal'' SNPs affecting phenotype, varying $\PVE$
from 0.01 to 0.5 in steps of 0.01.
Our Bayesian model assumes, through the prior on $\bfb$, that the
effect size distribution $\mathcal{E}$ is normal.
To check for robustness to deviations from this assumption, we
simulated phenotype data using both $\mathcal{E}= N(0,1)$ (as
effectively assumed by our model)
and $\mathcal{E} = \DE(1)$, where $\DE$ denotes the double
exponential distribution.
The results from these two different distributions were qualitatively
similar, and so we show only the results for $\mathcal{E} = \DE(1)$.

\begin{figure}
\centering
\begin{tabular}{cc}

\includegraphics{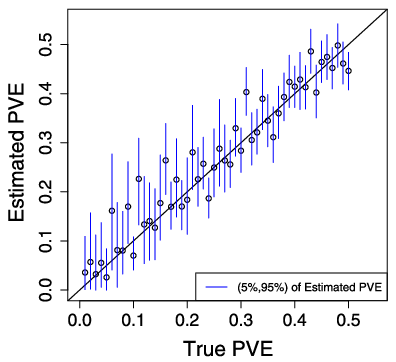}
&{\includegraphics{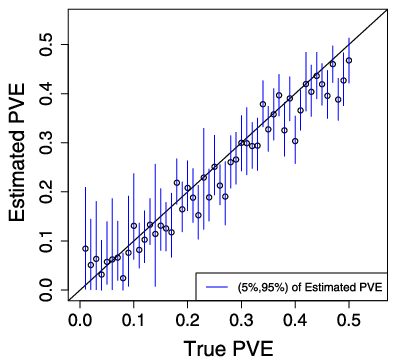}}\\
\footnotesize{(a)}&\footnotesize{(b)}
\end{tabular}
\vspace*{-3pt}
\caption{Comparison of true and inferred values for the proportion of
variance in $\bfy$ explained by relevant covariates $(\PVE)$. Panel
 \textup{(a)} shows results for $1,000$ individuals with $10,000$
independent simulated SNPs; Panel  \textup{(b)} shows results
for $980$ individuals with $317$K real SNP genotypes. Circles indicate
posterior mean for $\PVE$; vertical bars indicate the symmetric $90\%$
credible interval.}
\label{fig:herit}
\end{figure}

Figure \ref{fig:herit}
shows estimates
of $\PVE$ obtained by our method against the true values. For both
simulated and real SNP data there is a generally good correspondence
between the true and inferred values, and 90\% credible intervals (CI)
for $\PVE$ covered the true value in 85\% of cases. As might be
expected, the uncertainty in $\PVE$ is greater when there is a larger
number of SNPs, presumably due to the increased difficulty in reliably
identifying relevant variants. In addition, the uncertainty in $\PVE$
tends to be greater when the true $\PVE$ is smaller. Our intuition is
that when the data contain no SNPs with strong individual effects, it
remains difficult to rule out the possibility that many SNPs may have
very small effects that combine to produce an appreciable $\PVE$.
Nonetheless, even when the true $\PVE$ is small, the inferred
posterior interval for $\PVE$ does exclude large values, illustrating
that even in this case our method is able to extract information from
the data.\looseness=-1

\subsection{Many causal SNPs with tiny effects}

The simulations above involve 30 causal SNPs explaining \textit{in total}
between 0.01 and 0.5 of the total variance in $\bfy$.
We note that this is a relatively subtle level of signal: in the
following sections we will see that, for $\PVE=0.30$, and the sample
sizes we used, it is typically not possible to confidently identify the
majority of causal SNPs, nor to achieve the predictive performance that
is similar to one would obtain if one knew the causal variants. Thus,
to estimate the $\PVE$, BVSR must not only identify variants that are
confidently associated with $\bfy$, but also
estimate how many additional variants of small effects it might be
missing and what their effect sizes might be.
Clearly, there must be some limit to its ability to accomplish all
these tasks: in particular, if there
were very many variants of minuscule effects, then it would be
difficult to distinguish this from the null model in which
no variants have any effect. To try to test
these limits, we ran more challenging simulations involving many more
SNPs with tiny individual effects, but a nontrivial overall $\PVE$.
Specifically, we considered two cases:
(i) 300 causal SNPs out of the $10$K simulated SNPs in $1\mbox{,}000$
individuals; and (ii) $1\mbox{,}000$ causal SNPs out of the $317$K real SNPs
in $980$ individuals. In each case we simulate the effect sizes using
a~normal distribution. We simulated $10$ independent sets of phenotypes
with $\PVE=0.3$ in each case. For comparison in each case we also
simulated 10 independent sets of phenotypes under a ``null'' model with
no causal SNPs ($\PVE=0$).

\begin{figure}
\centering
\begin{tabular}{@{}c@{\hspace*{3pt}}c@{}}

\includegraphics{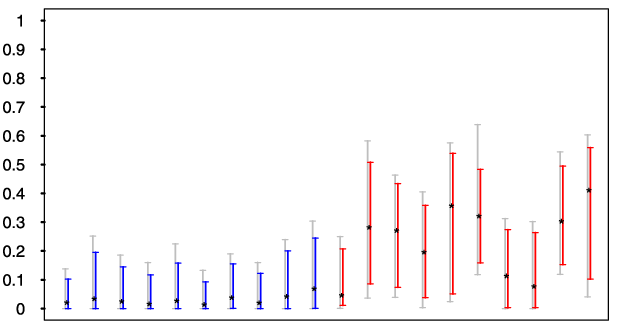}
&{\includegraphics{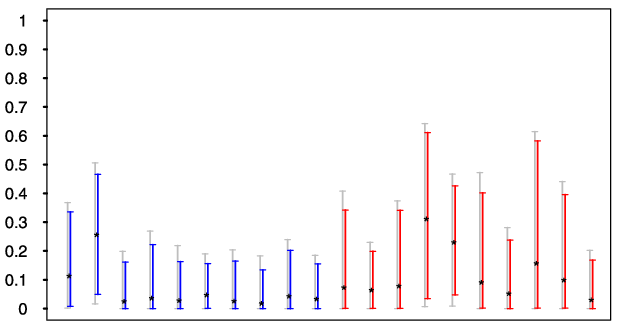}}\\
\footnotesize{(a)}&\footnotesize{(b)}
\end{tabular}%
\vspace*{-3pt}
\caption{Plots showing estimation of $\PVE$ for simulations with
large numbers of causal variants of very small effect.
Panels  \textup{(a)} and  \textup{(b)} are for $10$K and
$317$K data sets, respectively.
The grey lines denote $99\%$ CI and colored lines denote $95\%$ CI. The
blue color indicates null simulation $(\PVE= 0)$, red indicates
alternative simulations $(\PVE=0.3)$. The * denotes the median.}
\label{fig:many}
\end{figure}

For these data sets, to give BVSR some chance to identify the large
number of causal SNPs, we increased $M$, the upper limit on the
expected number of nonzero regression coefficients in our prior on $\pi
$, to $M=1\mbox{,}000$.
Plots of 99\% and 95\% credible intervals for $\PVE$ in each
simulation are shown in Figure \ref{fig:many}.

Somewhat surprisingly, for the first set of simulations, with $300$
causal SNPs out of $10$K and $\PVE=0.3$, BVSR remains able to provide
reasonable estimates of $\PVE$: for example, for $5$ of the $10$
simulations the interquartile range of the posterior on $\PVE$ spans
the true value of $\PVE=0.3$, and in $7$ simulations the 90\%
symmetric CI includes $\PVE=0.3$.
Further, there is a~clear qualitative difference between the results of
$\PVE=0.3$ and $\PVE=0$.
Less surprisingly, for the extremely challenging case of $1\mbox{,}000$ causal
SNPs out of $317$K, the estimates of $\PVE$
are considerably less precise. However, even here, these admittedly
limited simulations appear
to show systematic differences between $\PVE=0.3$ and $\PVE=0$.
For example, for $\PVE=0.3$, $8$ of the 90\% CIs cover $\PVE=0.2$ and
$6$ CIs cover $\PVE=0.3$; whereas for $\PVE=0,$
only $2$ of the 90\% CIs cover $0.2$ and $1$ CI covers $0.3$.

\subsection{Identifying the causal SNPs}

In existing GWAS the vast majority of studies published so far restrict
their analysis to the simplest possible approach of testing each SNP,
one at a time, for association with phenotype. One possible advantage
of a multi-SNP analysis like ours is to improve power compared with
this simple single-SNP approach. However, since each SNP is typically
correlated with only a small number of other (nearby) SNPs, and so any
two randomly chosen SNPs will be typically uncorrelated, the gain in
power might be expected to be small (at least in the absence of
interactions among SNPs). Further, one might be concerned that if our
MCMC scheme does not mix adequately, then the results of the multi-SNP
approach could actually be worse than those from a simpler analysis.

We performed two types of simulations to investigate these issues, the
first using the
10K data set (independent SNPs), and the second using the chromosome 22
of the 550K data set ($9\mbox{,}041$ correlated SNPs).
In each case we simulated $100$ phenotype data sets as described above,
with $30$ causal SNPs and $\PVE=0.25$.

For the 10K simulations we compared BVSR, single-SNP analyses, and
LASSO in their ability to identify the causal SNPs
as follows. For BVSR and single-SNP analyses we first computed, for
each SNP, a measure of the evidence for association with phenotype. For
BVSR we used the PIPs [equation (\ref{post:inclusion})]; for
single-SNP analysis we used the univariate Bayes Factor as described in
Section \ref{subsec:other}. We then consider thresholding this measure
of evidence: for any given threshold,
we consider all causal SNPs exceeding the threshold to be true
positives, and all other SNPs exceeding the cutoff to be false
positives. We compare methods by constructing curves showing the
trade-off between true positives and false positives as the threshold
is varied.
For LASSO, we first computed the solution path as $\lambda$ varies.
Then, for each solution on this path we defined all causal SNPs with
nonzero regression coefficients to be true
positives, and all other SNPs with nonzero regression coefficients to
be false positives. We then
constructed curves showing the trade-off between true positives and
false positives as $\lambda$ is varied.

For the real (correlated) SNPs we performed a similar comparison, but
assessed the methods in their ability
to identify the correct genomic \textit{regions} rather than individual
SNPs. This is because the three methods
differ qualitatively in the way they identify SNP associations when
SNPs are correlated with one another:
single-SNP analyses tend to identify significant associations
at any SNP that is strongly correlated with a causal SNP; LASSO tends
instead to select just one or a few
correlated SNPs; and BVSR tends to spread the association signal (the
PIPs) out among correlated SNPs. While
it may be important to be aware of these qualitative differences when
interpreting results from the methods, they
are not our main interest here, and we assess the methods
at the level of regions in an attempt to reduce the influence of these
qualitative differences. (Further, it could be argued
that identifying regions of interest is the primary goal of GWAS.)
To describe the approach in more detail, we partitioned chromosome 22
into $200$ kb nonoverlapping regions (different choices of region size
that we tried produced
qualitatively similar results). We then used each method to assign each
region a ``region statistic'' indicating the strength
of the evidence for an association in that region.
For single SNP analysis we used the maximum single SNP Bayes factor
within each region; for BVSR we used the
sum of the PIP for SNPs in the region; and for LASSO we used the
penalty~$\lambda$ at which any SNP in that region is included in the
model. Similar to the SNP-level comparisons, we plot how true and false
positive regions vary as the threshold on the region statistic is
varied. (We averaged results over
two different starting positions for the first window, $0,$ and $100$ kb.)

\begin{figure}
\centering
\begin{tabular}{cc}

\includegraphics{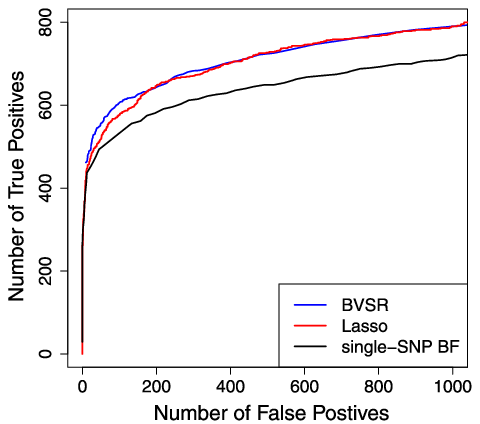}
&{\includegraphics{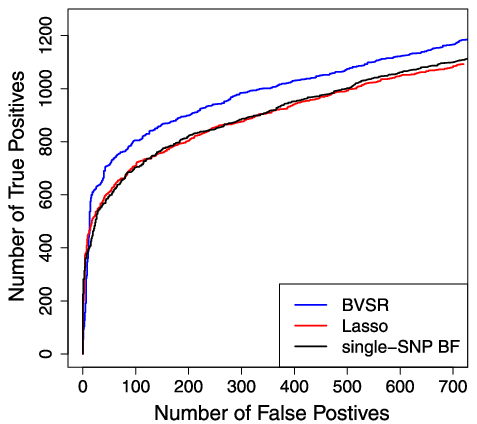}}\\
\footnotesize{(a)}&\footnotesize{(b)}
\end{tabular}%
\vspace*{-3pt}
\caption{Graphs showing the trade-off between true positive and false
positive SNP identifications for the different methods: BVSR $($blue$)$,
LASSO $($red$)$, and single SNP analyses $($black$)$. Both plots show results
that are summed across $100$ data sets $($see text for further
explanation$)$. Panel  \textup{(a)} is for independent simulated
SNPs; Panel  \textup{(b)} is based on real genotype data for
chromosome $22$.}
\label{fig:power}
\end{figure}

Figure \ref{fig:power}
shows curves of the trade-off between true and false positives for each
method in the two different simulations.
Each point on the curve shows the total true vs false positives across
the hundred simulated data sets,
using a common threshold across data sets. (An alternative way to
combine data sets is to use a different threshold in each data set,
vary the thresholds in such a way as to produce the same number of
positive findings in each data set;
the two different ways to combine data sets give similar results.)

For a given number of false positives, the multi-SNP approaches (BVSR
and LASSO) always yield as many or more true positives than the
single-SNP analysis. For the 10K simulated SNPs BVSR and LASSO perform
similarly, whereas for the
real genotypes BVSR is better. (The reasons for this difference are
unclear to us.) The results demonstrate that, even in the case where
single-SNP tests might be expected to perform extremely well---that
is, independent SNPs with no interactions---it is still possible to
gain slightly in power by performing multi-SNP analyses. Our intuitive
explanation for the gain in power of the multi-SNP approaches is that,
once one identifies a causal variant,
controlling for it will improve power to detect subsequent causal
variants. Because the SNPs are independent, this gain is expected to be
small: indeed, if the SNPs were exactly orthogonal, then one would
expect no gain by controlling for identified variants. However, our
results show that even in the case of independent SNPs the gain is
measurable because the finite sample size produces nonzero sample
correlations between ``independent'' SNPs.

We note that, at least in these simulations, most of the gain from the
multi-SNP methods occurs when the number of false positives is small
but nontrivial: that is, the multi-SNP methods promote some of the
moderately-difficult-to-detect causal SNPs slightly higher in the SNP
rankings, but not so far as to put them at the very top. This suggests
that multi-SNP analysis may be most useful when
used in combination with other types of data or analysis
that attempt to distinguish true and false positives among the SNPs
near the top of the association rankings [as in \citet
{raychaudhuri09}, e.g., where
information on gene similarities taken from PubMed abstracts are used
in this way].

\subsection{Prediction performance}

We used the same simulated data as in the previous section to compare
predictive performance of BVSR and LASSO.
To measure predictive accuracy, we use the relative prediction gain,
defined at (\ref{eqn:rpv}).
For our method we compute $\RPV(\bar{\bfb})$ where $\bar{\bfb}$
is the posterior mean for $\bfb$. For LASSO we compute the $\RPV$ in
two ways,
which we will refer\vspace*{1pt} to as $\RPV_1$ and $\RPV_2$.
For $\RPV_1$ we first compute $\RPV(\bfb^{(i)})$ for
each $\bfb^{(i)}$ in the solution path
for $\bfb$ output by the \texttt{lars} package, and take the minimum of these
relative prediction errors. Note that by taking
the minimum over~$\lambda$ in this way we are effectively assuming
that an
oracle has given us the optimal value for $\lambda$; in practice, one
would need
to obtain $\lambda$ through other means, such as cross-validation, which
would result in worse accuracy than $\RPV_1$.
For $\RPV_2$ we take a two-stage approach to prediction. First, we use LASSO
to select the SNPs that should have nonzero coefficients (using the
$\lambda$ used for $\RPV_1$),
and then we estimate the regression coefficients of these SNPs using
ordinary least squares ($\bfb_{{\mathrm{OLS}}}$), and compute $\RPV
_2:=\RPV(\bfb_{\mathrm{OLS}})$.
The motivation for this procedure is that if LASSO is able to reliably
identify the correct coefficients,
then the refitting procedure
will improve predictive performance by avoiding the known tendency for
LASSO to overshrink nonzero regression coefficients; however, as we
shall see below,
the refitting can be counterproductive when the correct coefficients
are not reliably identified.

\begin{figure}[b]

\includegraphics{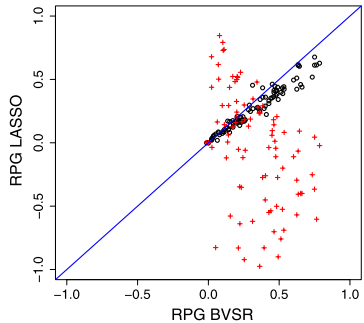}

\caption{Comparison of the relative prediction gain $(\RPV)$ for BVSR
$($x-axis$)$ and LASSO $($y-axis$)$.
Black circles are results from the optimal LASSO solution without
refitting $(\RPV_1)$; Red crosses are corresponding results with
refitting $(\RPV_2)$, described in the main text.}
\label{fig:rpv}
\end{figure}

Figure \ref{fig:rpv} compares the $\RPV$ obtained from the three methods
on 100 simulated data sets. The $\RPV$ from our Bayesian approach is
higher than that obtained directly from the optimal LASSO\vadjust{\eject} solution
($\RPV_1$) in $82$ of the $100$ data sets, and mean $\RPV$ is higher
($0.315$ vs $0.261$). The refitting procedure has a substantial effect
on predictive accuracy, and, in particular, it substantially increases the
variance of the performance: for some data sets the refitting procedure
improves predictive performance, but for the majority of data sets it
results in
much worse $\RPV$. Indeed, $\RPV_2$ is often negative, indicating
that predictive performance after refitting is substantially worse than
simply using the mean
phenotype value, which is the symptom of ``overfitting.'' This behavior
makes intuitive sense: in cases when the~optimal LASSO solution does a
good job of precisely identifying many of the relevant covariates, and
no irrelevant ones, the refitting step improves predictive performance,
but when the first stage includes several
false positives the refitting procedure is counter-productive.

Although our Bayesian model is sparse, our estimated $\bar{\bfb}$ is
not sparse due to the averaging in (\ref{eqn:betarb}). In some
contexts one might want
to obtain a~sparse predictor, so, to examine how this might impact predictive
accuracy, we computed the $\RPV$ for each data set using only the $P$
covariates with highest posterior inclusion probabilities (setting
other coordinates of $\bar{\bfb}$ to~0), where $P=10, 30, 100$. The
average $\RPV$ for
these sparse estimates of $\bfb$ were essentially unchanged from using
the nonsparse estimate $\bar{\bfb}$ ($\RPV= 0.313, 0.315,$ and
$0.315$, resp.).

We also examined the benefits of using Bayesian model averaging (BMA)
to perform prediction, by computing the $\RPV$
obtained using only those covariates with a posterior inclusion
probability ${>}t$ where $t=0.2,0.5,0.8$. [When $t=0.5$ this is the
``median probability model'' of
\citet{barbieriberger04}.]
Specifically, we computed the $\RPV$ for $\hat{\bfb}_j = I(\hat{\Pr
}(\gamma_j =1)> t) \hat{E}(\beta_j |\break \gamma_j=1)$, where the two
quantities on the right-hand side are estimated
from~(\ref{post:inclusion}) and (\ref{eqn:betarb}). These estimates
have some shrinkage because $E(\beta_j | \gamma_j=1)$ is a~shrinkage
estimate of $\beta_j$ (due to the normal prior on $\bfb$), but they
do not have the additional shrinkage term $\Pr(\gamma_j =1)$ that BMA
provides to further shrink variables that are not confidently included
in the model.
The average $\RPV$'s for these non-BMA estimates were notably worse
than for the BMA-based estimates: $0.244, 0.291,$ and $0.272$,
respectively, compared with 0.315 for BMA.

Taken together, these results suggest that BMA is responsible for a
moderate amount of the gain in predictive performance of BVSR compared
with $\RPV_1$,
with some of the remainder being due to LASSO's tendency to over-shrink
estimates of the nonzero regression coefficients. One way to think of
this is that
LASSO has only a single parameter, $\lambda$, that controls both
shrinkage and sparsity. In this setting the true solution
is very sparse, so~$\lambda$ needs to be big enough to keep the
solution sufficiently sparse; but having $\lambda$ this big
also creates an overly strong shrinkage effect. In contrast, BVSR
effectively avoids this problem by having two
parameters, $\sigma_a$ controlling shrinkage, and $\pi$ controlling sparsity.
As we have seen, in this context the strategy of refitting the $\beta$
coefficients at the LASSO solution fails to improve average predictive
performance.
Other possible ways around this problem include using a more flexible
penalized regression model (e.g., the Elastic Net [\citet{zouhastie05}]
has two parameters, rather than one),
or using a procedure that does not overshrink large effect sizes, for
example, SCAD [\citet{scad}]. Comparisons of these methods with BVSR would
be an interesting area for future work.

\begin{figure}[b]
\centering
\begin{tabular}{@{}c@{\hspace*{4pt}}c@{\hspace*{4pt}}c@{}}

\includegraphics{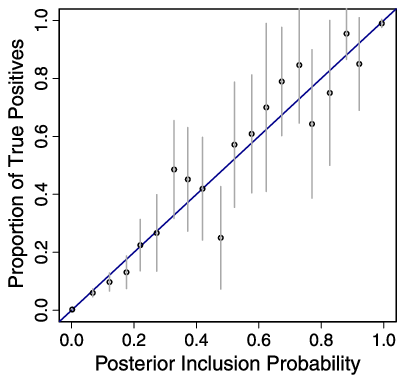}
&{\includegraphics{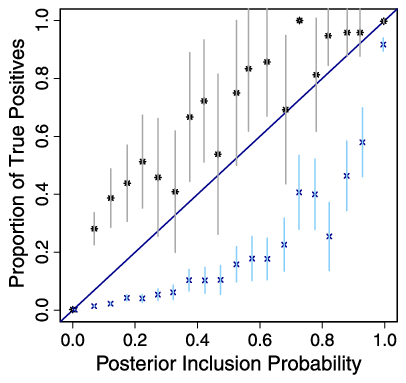}}&{\includegraphics{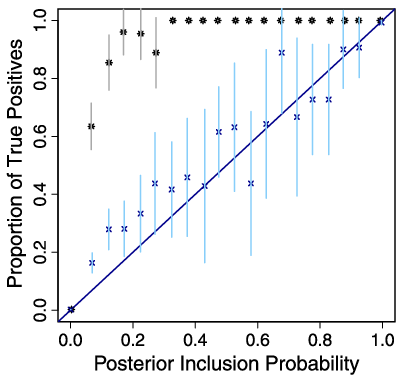}}\\
\footnotesize{(a)}&\footnotesize{(b)}&\footnotesize{(c)}
\end{tabular}%
\vspace*{-3pt}
\caption{Calibration of the posterior inclusion probabilities
$($PIPs$)$ from BVSR. The graph was obtained by binning the PIPs obtained from
BVSR in $20$ bins of width $0.05$. Each point on the graph represents a
single bin, with the $x$ coordinate
being the mean of the PIPs within that bin, and the $y$ coordinate
being the proportion of SNPs in that bin that were true
positives $($i.e., causal SNPs in our simulations$)$. Vertical bars show
$\pm 2$ standard errors of the proportions, computed from
a binomial distribution.
Panel  \textup{(a)} is the result of BVSR, using the priors
described here. The fact that the points lie near the line $y=x$
indicates that the PIPs are reasonably well calibrated, and thus
provide a reliable assessment of the confidence that each SNP should be
included in the regression. Panel  \textup{(b)} is the result
from BVSR fixing $\pi$ to be either $5\times$ smaller $($black star$)$ or
$5\times$ larger $($blue cross$)$ than the true value $(\sigma_a$ fixed
to true value$)$. Panel  \textup{(c)} is the result of fixing
$\sigma_a$ to be either $5\times$ smaller $($black star$)$ or $5\times$
larger $($blue cross$)$ than the true value $(\pi$ fixed to true value$)$.}
\label{fig:calibration}
\end{figure}

\subsection{Calibration of the posterior inclusion probabilities}

One of the main advantages of BVSR compared with Bayesian single-SNP
analysis methods is that BVSR allows
the hyperparameters $\pi$ and $\sigma_a$ to be estimated from the
data, and thus provides data-driven estimates
of the posterior inclusion probabilities (PIPs). One hope is that
estimating these parameters from the data will lead to
better-calibrated estimates of the PIPs than the single-SNP approach
which effectively requires one to supply educated guesses for these parameters.
To assess this, Figure \ref{fig:calibration}(a) shows the calibration of
the PIPs from BVSR, for the simulations used in the estimation of
$\PVE$ above (fifty data sets with $\PVE=0.01$--$0.5$ for both normal
and exponential effect size distributions). The figure shows that the
PIPs are reasonably well calibrated. In particular, SNPs with high PIP
have a high probability of being causal variants in the simulations.

To illustrate the potential benefits of using moderately-diffuse prior
distributions on $\pi$ and $\sigma_a$, allowing their values to be
informed by the data,
rather than fixing them to specific values, we also applied BVSR with
either~$\pi$ or~$\sigma_a$ fixed to an ``incorrect'' value
(approximately 5 times larger or smaller than the values used in the
simulations). Figure \ref{fig:calibration}(b) and  (c)
show how, as might be expected, this can result in poorly-calibrated
estimates of the PIP (of course, if one were lucky enough to fix both
$\pi$ and $\sigma_a$ to their ``correct'' values, then calibration of
PIPs will be good,
but, in practice, the correct values are not known). We note that
fixing $\sigma_a$ to be five-fold too large seems to have only
a limited detrimental effect on calibration, which is consistent with
the fact that in single-SNP analyses, with moderate sample sizes, BFs
are relatively insensitive to choice of $\sigma_a$ provided it is not
too small [e.g., \citet{stephensbalding09}, Figure 1]. This suggests
that, in specifying
priors on $\sigma_a$, it may be prudent to err on the side of using
a~distribution with too long a tail rather than too short a tail.
Note that, as in Bayesian single-SNP analyses, although the numerical
value of the PIP is sensitive to choice of $\pi$, the ranking of SNPs is
relatively insensitive to choice of $\pi$ (and, indeed, $\sigma_a$).
Consequently, in contrast to the calibration plot, power plots of the
kind shown in
Figure \ref{fig:power}
are not sensitive to choice of prior on either~$\pi$ or $\sigma_a$
(results not shown).

\subsection{Real data analysis: PARC GWAS for C-reactive protein}
\label{sec:parc}

We applied BVSR to analyze a GWAS study to identify genetic variants
associated with
plasma C-reactive protein (CRP) concentration. CRP is a protein found
in the blood that is
associated with inflammation, and is predictive of future
cardiovascular disease [\citet{ridkeretal02}].
The data come from the Pharmocogenetics and Risk of Cardiovascular
Disease (PARC) study [\citet{reineretal08} and references therein].

The available genotype data consisted of 1968 individuals genotyped on
either the Illumina 317K chip (980 individuals) or the Illumina 610K
SNP chip plus a custom 13,680 SNP Illumina i-Select chip (988
individuals). These genotype data
had undergone basic quality control filters (e.g., removing SNPs with
very high proportions of missing data, or showing strong departures
from Hardy--Weinberg equilibrium)
prior to our analysis. To merge the two data sets, we used genotype
imputation [\citet{servinstephens07}; \citet{marchinietal07}],
using the software package BIMBAM [\citet{guanstephens08}] to replace
missing or unmeasured genotypes with their posterior mean given
the observed genotype data [see \citet{guanstephens08} for discussion
of this strategy].
After imputing missing genotypes, we removed SNPs with (estimated)
minor allele frequency $< $0.01, leaving a~total of 530,691 SNPs.

The phenotype data consisted of plasma concentrations of CRP, measured
multiple times for each individual, both before and after exposure to
statin drugs.
These multiple\vadjust{\eject} measures were adjusted for covariates (age, sex, smoking
status, and body mass index), quantile normalized to a standard normal
distribution, and averaged to produce
a single summary measure of CRP concentration for each individual
(relative to other individuals in the same study), as described in
\citet{reineretal08}.

\begin{figure}
\centering
\begin{tabular}{c}

\includegraphics{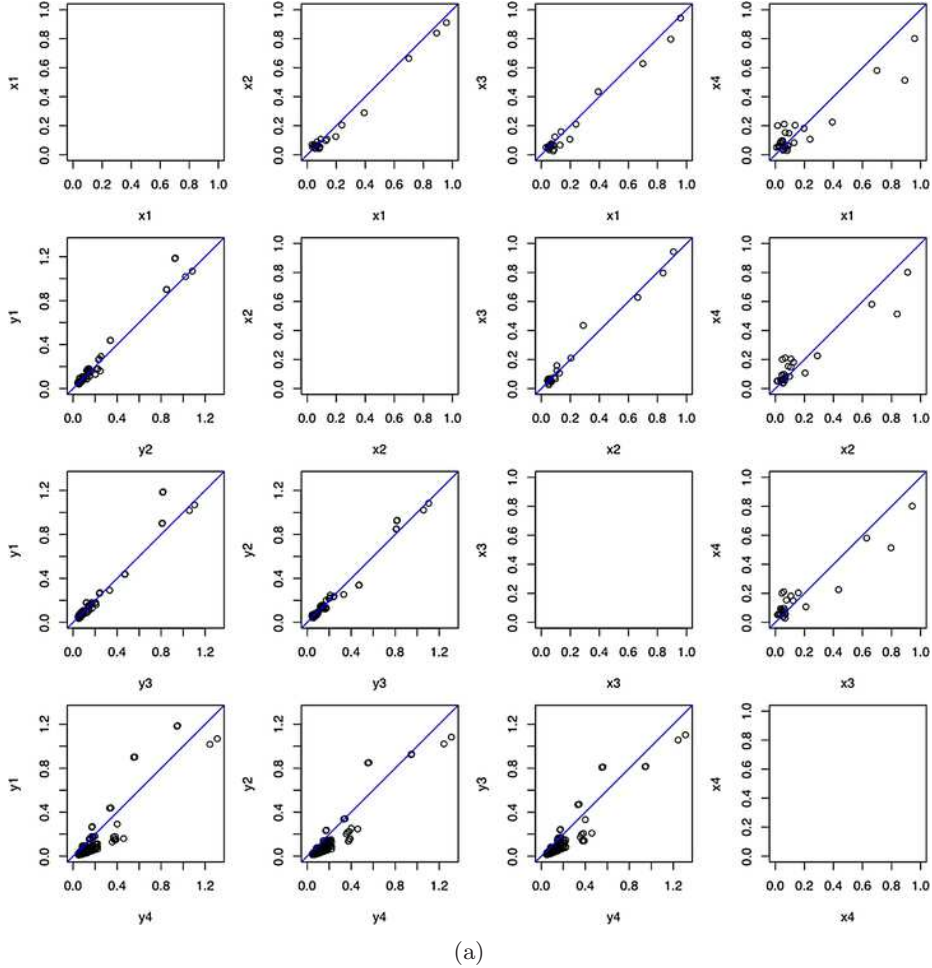}
 \\
\footnotesize{(a)}
\end{tabular}%
\vspace*{-3pt}
\caption{Illustration of the consistency of results across four
different runs of the MCMC algorithm for the CRP data. In panel
\textup{(a)} the $(i,j)$th plot compares results for runs $i$ and~$j$.
Plots in the upper triangle $(j>i)$ compare estimated posterior
inclusion probabilities $($PIPs$)$ for each SNP. Plots in the lower
triangle compare estimated posterior expected number of SNPs in 1 Mb
regions $($so each point corresponds to a single region$)$. The line $y=x$
is marked in blue. Panel  \textup{(b)} shows posterior
distributions of $\PVE$ from the four MCMC runs.}
\label{fig:crp}
\end{figure}

After removing individuals with missing phenotypes, we had phenotype
and genotype data on a total of $1\mbox{,}682$ individuals.
We performed four independent MCMC runs, two with $2$ million
iterations, and two using $4$ million iterations. These longer runs
took approximately 60 and 90 CPU hours on a~Mac Pro $3$ GHz desktop.
Comparing results among runs, we found three of the runs gave very good
agreement in all aspects we examined,
whereas the fourth run showed mild but noticeably greater departure
from the others, suggesting possible convergence or
mixing issues. For example, Figure \ref{fig:crp}(a) compares the
estimated PIPs for each pair of runs, and Figure \ref{fig:crp}(b)
compares the estimated posterior distribution of $\PVE$ among runs.
The remainder of the results in this section are based on pooling the
results from all four runs.

\setcounter{figure}{5}
\begin{figure}
\centering
\begin{tabular}{c}
{
\includegraphics{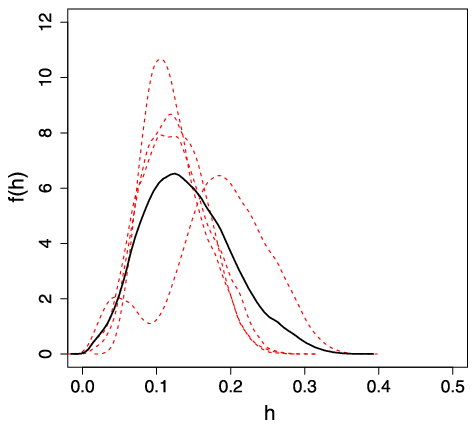}
}\\
\footnotesize{(b)}
\end{tabular}%
\vspace*{-3pt}
\caption{$($Continued$)$.}
\end{figure}

The usual way to summarize single-SNP analyses is to report the SNPs
with the strongest marginal
evidence for association. Thus, it might seem natural in a multi-SNP
analysis to focus on the SNPs with the largest
posterior inclusion probabilities (PIPs). However, this can be
misleading. For example, if there are many SNPs in a region
that are highly correlated with one another, and all approximately
equally associated with the phenotype, then
it may be that the correct conclusion is that at least one of these
SNPs should be included in the model, but there
might be considerable uncertainty about which one. In this case, even
though the posterior probability of at least one SNP
being included in the model would be high (near~1), none
of the individual PIPs may be very big, and concentrating on the PIPs
alone would risk missing this signal in the data.
To avoid this problem, we prefer to initially summarize results at the
level of \textit{regions}, as we now illustrate.

We divided the genome into overlapping regions, each 1 Megabase ($10^6$
bases) in length, with the overlap between adjacent regions being 0.5
Megaba\-ses. For each region we computed two quantities: (i) an estimate,
$E$, of the posterior expected number of SNPs included in the model,
being the sum of the estimated PIPs for all SNPs in the region; (ii) an
estimate of the probabilities, $P$, that the region contains (a) 1 SNP,
(b) 2 SNPs, or (c) more than 2 SNPs included in the model.
The latter quantities (ii) are perhaps the most natural summary of the
evidence that the region harbors genetic variants affecting phenotype,
but (i) has the advantage that it can be easily approximated using
Rao--Blackwellization, resulting in lower Monte Carlo error. Thus, in
practice, we suggest
examining both quantities, and placing more trust in (i) where the two disagree.

\begin{figure}
\centering
\begin{tabular}{c}

\includegraphics{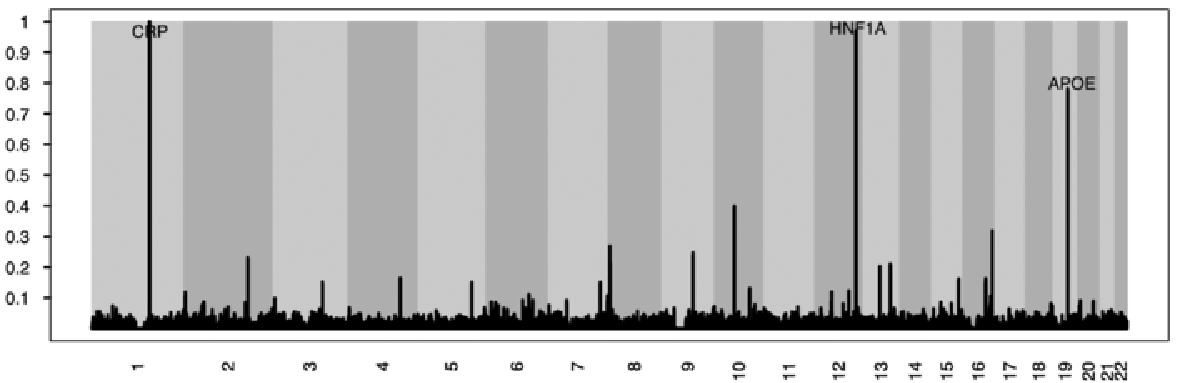}
\\
\footnotesize{(a)}\\[6pt]

\includegraphics{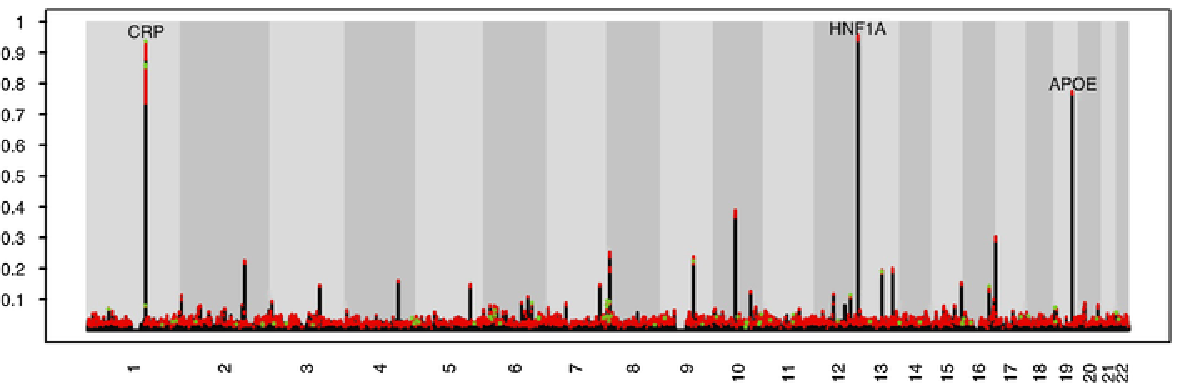}
\\
\footnotesize{(b)}\\[6pt]

\includegraphics{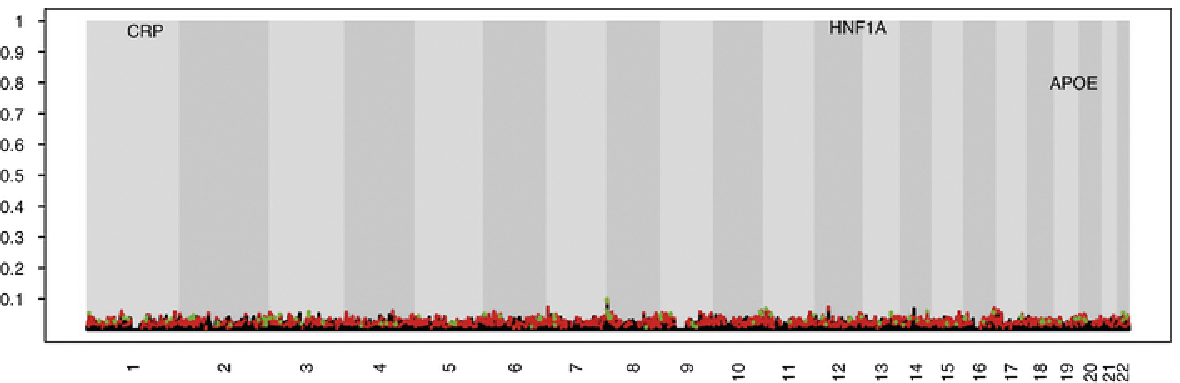}
\\
\footnotesize{(c)}
\end{tabular}%
\vspace*{-3pt}
\caption{For each $1$ Mb region we show an estimate from BVSR that the
region contains $1$ $($black$)$, $2$~$($red$)$, or more than $2$ $($green with $\star
)$ SNPs in the regression. The $1$ Mb regions overlap by $0.5$~Mb, and so any
SNP with a large PIP would cause a signal to occur in $2$ adjacent
regions on the plot. Panel  \textup{(a)} shows sum of PIP in
each $1$ Mb region $($truncated at $1)$.
Panel  \textup{(b)} shows estimated probabilities that each
genomic region harbors variants associated with CRP levels. In panel
 \textup{(c)} we permute phenotype once and produce the same
plot as a comparison.}
\label{fig:crppost}
\vspace*{3pt}
\end{figure}

The results are summarized in Figure \ref{fig:crppost},
which also shows results for a single permutation of the phenotypes for
comparison.
The plot clearly identifies two regions with very strong evidence for
an association with CRP in both plots (e.g., $E>0.95$), and a third region
with moderately strong evidence (e.g., $E>0.75$). Multiple other
regions show modest signals ($E=0.1$ to 0.5), that might generally be
considered worthy of follow-up in larger samples, although at this
level of signal the majority are, of course, unlikely to be truly
associated with CRP.

The three regions with the strongest association signals contain the
genes   \textit{CRP}, \textit{HNF1A},  and \textit{APOE/APOC}, all of which have shown
robustly-replica\-ted SNP associations with C-reactive protein levels in
several other GWAS using single-SNP analyses [e.g., \citet
{reineretal08}; \citet{ridkeretal08}]. In addition, in these data, these
three regions all contain single SNPs showing strong associations: the
largest single-SNP Bayes factors in each of these regions are
$10^{6.2}, 10^{5.5}$, and $10^{4.9}$, respectively.
Thus, in this case the identification of regions of interest from BVSR
is largely concordant with what one would have obtained from a
single-SNP analysis. However, we highlight two advantages of the BVSR
analysis. First, the estimated posterior probabilities obtained for
each region are easier to interpret
than the single-SNP Bayes Factors. For example, the estimated posterior
probability that the HNF1A region contains at least one SNP included in
the model is $0.96$, and this seems much more helpful than knowing that
the largest single-SNP Bayes Factor in the region is $10^{4.9}$.
Similarly, for the next most associated region, which is on chromosome
10 near the gene FAM13C1, the posterior probability
of $0.42$ is simpler to interpret than the fact that the largest
single-SNP Bayes Factor is $10^{3.9}$. And while these single-SNP BFs
are easily converted to
posterior probabilities of association by specifying a prior
probability of association (effectively $\pi$ in our model), the
multi-SNP analysis reduces the risk of specifying an inappropriate
value for $\pi$ by learning about $\pi$ from the data.\looseness=-1

A second advantage of BVSR is its ability to estimate the PVE.
To illustrate this, we first consider a typical single-SNP analysis in
this context,
which estimates the PVE for ``significant'' SNPs by performing ordinary
least-squares
regression on those SNPs. Applying this approach to these data, using a
relatively liberal
(by GWAS standards) threshold for significance (single-SNP BF $>$10$^4$),
we find that significant SNPs explain approximately $6\%$ of the
overall variance in CRP after controlling for covariates.
Comparing this with some previous estimates of heritability of CRP in
the range 0.35--0.4 [\citet{pankow2001}; \citet{lange2006}]
suggests that a substantial amount of genetic variation influencing CRP
remains to be identified, a feature that
has become known as ``missing heritability.''
One question of interest is to what extent this shortfall might be
explained by measured genetic variants
that simply failed to pass the significance threshold, vs being
explained by unmeasured genetic variants or other factors.
To assess this, we examine $\PVE$ obtained from applying BVSR on
measured SNPs.
The posterior distribution for PVE [Figure \ref{fig:crp}(b)] has mean
$0.14$, with a symmetric 90\% CI of $[0.05, 0.25]$. Note that, as one
might expect,
the lower part of this CI is similar to the estimated PVE of
``significant'' SNPs.
Because most of the posterior distribution
lies above 0.06, we infer that larger studies of \textit{the same set of SNPs}
might be expected to uncover considerably more signal than this study.
(Consistent with this,
a larger study involving 6,345 women typed at a subset of the SNPs
considered here identified four additional genome regions containing SNPs
associated with CRP levels [\citet{ridkeretal08}].)
Conversely, the fact that the upper part of the CI (0.25) remains
well short of previous estimates of heritability suggests
that not all of the missing heritability is likely to be explained by
simply conducting larger studies of the same SNPs, and that
some alternative factors (e.g., unmeasured rare variants) may also contribute.

\section{Extension to binary phenotypes}
\label{sec:binary}
Although we have focused here on quantitative traits, BVSR is also
potentially applicable to binary phenotypes,
and this is important for GWAS applications because they often involve
binary phenotypes.
In this section we briefly summarize our attempts to extend BVSR in
this way.

A standard approach to applying BVSR to binary phenotypes is to use
a~probit link function. In practice, this is usually accomplished
by introducing latent variables $\bfz$ which are assumed to follow the
standard linear regression (\ref{model}), and to
be related to the observed outcomes $\bfy$ by $y_i = 1$ $(z_i>0)$ [\citet
{albertchib93}]. Posterior inference is performed
by integrating out $\bfz$ using Markov chain Monte Carlo, which
requires implementation of only one additional
update compared with the quantitative trait (an update of the $\bfz$
variables).

A nice feature of
this probit-based approach using latent Gaussian variables is that it
would allow us to use the same priors as for quantitative
outcomes, except that these priors now relate to the
unobserved latent (Gaussian) variables, rather than the observed
(binary) outcomes. Furthermore, we can
continue to summarize the overall signal by estimating the $\PVE$ of
the latent variables.
However, the way we have set things up, with an improper prior
on $\tau$, this would lead to improper posteriors on $\tau$ and $\bfz$
[because the likelihood $p(\bfy| \bfz)$ is unchanged
by multiplying $\bfz$ by any positive constant]. This could be
rectified in a number of ways.
For example, we could fix $\tau$ [e.g., to 1, as in \citet{albertchib93}].
Here we instead choose to impose an identifiability constraint directly
on the elements of $\bfz$, by constraining them
to have (empirical) variance 1, because this
allows us to re-use exactly the same computer code as for the
quantitative phenotypes
(whereas fixing~$\tau$ would necessitate some changes).
In addition, in an attempt to improve mixing, we make the approximation
that the marginal
distribution of the elements of $\bfz$ will be normal, which should be
a reasonable approximation
under the linear regression model (\ref{model}) provided that there
are no very large values for $\beta$.
Specifically, we restrict $z_1,\ldots ,z_n | \bfy$
to take a fixed set of values, being the $n$ equally-spaced quantiles
of a standard normal $N(0,1)$ distribution,
with the values corresponding to the $n_0$ individuals with $y_i=0$
being constrained to be the first
$n_0$ of these quantiles. The intuitive motivation for this constraint is
that it can reduce the potential to fall into poor local optima by
ruling out
implausible configurations of $\bfz$ that correspond to some SNPs
having very large effects. (Of course,
this may not be a good idea in settings where very large effects are
more plausible.)
With this constraint in place, local Metropolis--Hastings proposals for
$\bfz$ simply involve randomly picking a pair of
individuals $(i,j)$ with the same (binary) phenotype value and
proposing to swap the values of $z_i$ and $z_j$.
(For long-range proposals, we simply compound this local proposal
randomly many times.)

\begin{figure}
\centering
\begin{tabular}{cc}

\includegraphics{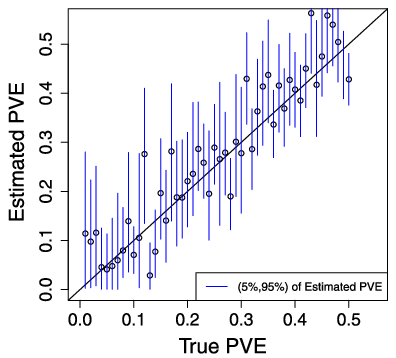}
&{\includegraphics{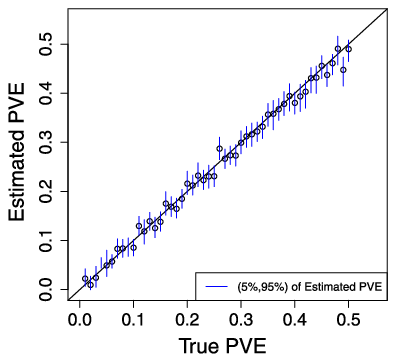}}\\
\footnotesize{(a)}&\footnotesize{(b)}
\end{tabular}%
\vspace*{-3pt}
\caption{Comparison of true and inferred values of $\PVE$ for binary
phenotypes. The estimated $\PVE$ is on the $y$-axis and the true $\PVE
$ on the $x$-axis. Panels  \textup{(a)} and  \textup{(b)} are for $n=1\mbox{,}000$ and $n=6\mbox{,}000$
individuals,
respectively. Circles indicate posterior mean for $\PVE$; vertical
bars indicate the symmetric $90\%$ credible interval.} \label{fig:binary}
\end{figure}

To provide a brief illustration of the potential for this approach,
we applied the method to some simple simulated data sets.
The genotypes were simulated in the same way described in Section \ref
{sec:sim}, using 10,000 independent
SNPs genotyped in $n=1\mbox{,}000$ and $6\mbox{,}000$ individuals. We simulated
latent normal phenotypes by randomly selecting $30$ causal SNPs and
simulating a quantitative phenotype $\bfz$ with prespecified $\PVE$
as in Section \ref{sec:sim}. We then converted these $n$ quantitative
phenotypes to $n$ binary phenotypes by mapping the largest $n/2$ $z$
values to $y=1$ and the remainder to $y=0$.
Figure \ref{fig:binary} illustrates how reliably we are able to infer
the PVE of the latent variables
from the binary data. More generally, we find that provided we limit
analyses to thousands of SNPs,
we are able to obtain generally reliable results for binary traits
(e.g., results from multiple independent runs largely agree with one another).
Thus, for example, we should be able to obtain reliable results for
small genomic regions, such as individual genes,
which can itself be of considerable interest [\citet
{servinstephens07}]. However, our experience with larger
real data sets involving hundreds of thousands of SNPs indicates that
mixing is, as one might expect, harder for binary traits than for
quantitative traits, and that to obtain reliable results
in practice for GWAS may require longer MCMC runs and/or further
methodological innovation.

\section{Discussion} \label{sec:discussion}

In this paper we have
demonstrated that BVSR can be successfully applied to large problems,
with a particular focus on
genome-wide association studies. We have argued that BVSR has several
potential benefits compared with
standard single-SNP analyses, among them the ability to obtain
data-driven estimates of hyperparameters that must
otherwise be specified more subjectively by the user, and the ability
to estimate the overall signal (PVE) that might
be accounted for by relevant covariates, even when confidently
identifying the relevant covariates is not possible.
We have also introduced a novel, more interpretable, approach to prior
specification
in BVSR, and shown that BVSR can provide a competitive alternative to the
penalized regression procedure LASSO.

However, despite our generally upbeat assessment, there are a number of
potential limitations of the methods
we have described here, which present both pitfalls to be aware of in
practice, as well as challenges and
opportunities for future work.

One important aspect of analysis of any GWAS is the potential for data
quality to adversely impact results. For example,
although modern genotyping technologies provided very high quality
genotypes on average, some SNPs are much harder
to genotype accurately than others, and genotyping error can occur at
some SNPs at an appreciable rate. This can cause
false positive associations if genotyping error is correlated with
phenotype (which it can be, particularly in case-control
studies if the DNA quality differs appreciably between cases and
controls [\citet{clayton05}]). While quality control is vital to any study,
it is of potentially even greater import in multi-SNP analyses than in
single-SNP analyses, because in multi-SNP analyses the association
results at one SNP affect the results at other SNPs, and so low quality
data at a few SNPs may impact estimated associations at other SNPs.
Thus, it seems particularly important to attempt to impose stringent
data quality
filters before embarking on a computationally-intensive multi-SNP analysis.

One limitation of the methods we present here is the assumption that
effect sizes are normally distributed. Although
our simulations with exponential effect sizes suggest a certain amount
of robustness to this assumption, it is important
to note that there are some phenotypes where the normality assumption
is clearly wrong. For example,
in type 1 diabetes, one region of the genome, the MHC, contains genetic
variants whose effect on phenotype may be substantially greater than
any other region. When such regions of unusually large effect are
known, it would be prudent to run methods like ours both including and
excluding data at these loci, to check for robustness of conclusions.
More generally, the robustness of our BVSR could be improved by
replacing the assumption of
normally-distributed effects with a heavy-tailed
distribution such as a $t$ with small or moderate degree of freedom, or
indeed with a prior on the degrees of freedom.

Another related issue is that we assume the residual distribution of
the phenotypes to be normal. To improve
robustness to this assumption, we typically normal quantile transform
the observed phenotypes to have a normal distribution before analysis
(which while not strictly ensuring that the residuals are normal, does
in our experience limit problems that might otherwise
be caused by deviations from normality, such as occasional outlying
values). Again, the use of a $t$ distribution
for the residuals might be preferable.\looseness=1

We view the work presented here as just the very start of what could be
done with BVSR in GWAS. One important extension would be
to incorporate additional information into the prior distribution on
which variables are included in the regression ($\bgamma$
in our notation). Here we have assumed that variables are included in
the model, independently, with common probability $\pi$.
This independence assumption ignores likely local spatial dependence of
$\bgamma$. In particular, it would be
unsurprising to see multiple functional variants occurring in a single
gene, and, indeed, analyses of genetic data in the CRP gene
have suggested that it contains multiple SNPs affecting CRP levels
[\citet{verzillietal08}; \citet{stephensbalding09}]. The independence
assumption in the prior we use here makes it overly skeptical about
this possibility. Another important possibility
is that one could allow the prior probability of each SNP being
included in the regression to depend on annotations of
the SNP, such as where it lies relative to a gene, or whether it lies
in a genomic region that is conserved across several species (a sign
that the region may be functional). Of course it is not generally known
  a priori  how much such annotations
should affect the prior inclusion probabilities. However, with BVSR one could
\textit{estimate} hyperparameters that affect the prior inclusion
probabilities from the data [\citet{veyrierasetal08}].

Finally, despite our focus on GWAS, many of the issues we have
discussed here have broad relevance. In particular, while
the computational challenges of BVSR remain considerably greater than
penalized regression methods, we believe that the
qualitative advantages of BVSR make it worth investing
effort into designing more efficient inference algorithms
for BVSR, to be able to better deal with the very large-scale
applications that are becoming increasingly common.

\begin{appendix}
\section{Details of MCMC scheme}\label{appmA}

We use Markov chain Monte Carlo to obtain samples from the posterior
distribution of $(h, \pi, \bgamma)$ on the product space of $(0,1)
\times(0, 1) \times\{0,1\}^p$, which is given by
\begin{equation} \label{post:hsg}
\pit(h, \pi, \bgamma| \bfy) \propto\pit(\bfy| h, \bgamma) \pit
(h) \pit(\bgamma| \pi) \pit(\pi).
\end{equation}
Here we are exploiting the fact that the parameters $\bfb$ and $\tau$
can be integrated out analytically to
compute the marginal likelihood $\pit(\bfy| h, \bgamma)$.
Indeed, in the limit for the hyperparameters $\lambda, \kappa
\rightarrow0$ and $\sigma_\mu\rightarrow\infty$
that we use here, we have
%
\begin{equation}\label{eqn:bf}
\qquad\frac{\pit(\bfy| h,\bgamma)}{\pit(\bfy| h, \bgamma=\mathbf{0})} =
n^{1/2} |\Omega|^{1/2} \frac{1}{\sigma_a(h,\bgamma)^{|\bgamma|}}
 \biggl( \frac{\bfy^t\bfy- \bfy^tX_\bgamma\Omega X_\bgamma^t \bfy
}{\bfy^t \bfy- n\bar{\bfy}^2}  \biggr)^{-n/2},
\end{equation}
where $\Omega:= (\sigma_a(h,\bgamma)^{-2} I_|\bgamma| + X_\bgamma
^t X_\bgamma)^{-1}$ and $\mathbf{0}$ denotes the $p$-vector of all 0s.
[For derivation, see \citet{servinstephens07}, Protocol S1 equation
(13).] Note that here $\sigma_a(h,\bgamma)$ is given by equation
(\ref{eqn:sigma}).

For each sampled values of $h,\bgamma$ from this posterior, we obtain
samples from the posterior distributions of $\bfb$ and $\tau$ by
sampling from their conditional distributions given $\bfy,\bgamma,h$:
%
\begin{eqnarray} \label{post:beta}
\tau| \bfy, h, \bgamma&\sim&\Gamma\bigl(n/2, 2/(\bfy^t \bfy- \bfy^t
X_\bgamma\Omega X_\bgamma^t \bfy)\bigr), \nonumber\\
\bfb_\bgamma| \tau, \bfy, h, \bgamma&\sim& N\bigl(\Omega X_\bgamma^t
\bfy, (1/\tau) \Omega\bigr) ,\\
\bfb_{-\bgamma} | \tau, \bfy, h, \bgamma&\sim&\delta_0.\nonumber
\end{eqnarray}

Our Markov chain Monte Carlo algorithm for sampling $h,\pi,\bgamma$
is based on a Metropolis--Hastings algorithm [\citet
{metropolisetal53}; \citet{hastings70}], using a simple local proposal to
jointly update $h, \pi,\bgamma$. In outline, the local proposal
proceeds as follows. First a new proposed value of $\bgamma$, $\bgamma
'$, is obtained by small modification of the current value (see below
for more details); then a~new value of $\pi$ is proposed from a
$\operatorname{Beta}(|\bgamma'|,p-|\bgamma'|+1)$ distribution; finally a proposed
new value for $h$ is obtained by adding a $U(-0.1,0.1)$ random variable
to the current value (reflecting proposed values that lie outside
$[0,1)$ about the boundary). The proposal distribution for $\pi$ is
proportional to its full conditional distribution given $\bgamma'$
inside the finite range of the prior on $\pi$ [given by (\ref
{prior:s})]; on the infrequent occasions that the proposed value for
$\pi$ lies outside this range, it is of course rejected.

In addition to the local proposal described above, we sometimes (with
probability 0.3 each iteration) make a longer-range proposal by
compounding randomly-many local moves [the number being uniform on
$(2,\ldots ,20)$]. This technique, named ``small-world proposal,'' improves
the theoretical convergence rate of the MCMC scheme [\citet{guankrone07}].

We now give details on our update proposal for $\bgamma$.
When adding a covariate into the model we use a \textit{rank based
proposal} that focuses more attention on
covariates that are more likely to be included in the model. To do
this, we first rank the covariates based on their association with
phenotype $\bfy$ (specifically we rank them by the Bayes factor for
the model including only that covariate vs the null model containing no
covariates, evaluated at $\sigma_a=1$). Let $Q_t$ be a distribution on
$(0, \ldots , t-1)$ which has decreasing probability. Here we choose
$Q_t$ to be a mixture $Q_t = 0.3 U_t + 0.7 G_t$, where $U_t$ is a
uniform distribution on $\{0,\ldots ,t-1\}$ and $G_t$ is a geometric
distribution truncated to $\{0,\ldots ,t-1\}$,
with its parameter chosen to give a mean of 2,000.

Now let $\bgamma^+$ denote the set of covariates that are currently in
the model, $\bgamma^+ = \{i\dvtx  \gamma_i = 1\}$. Let $\bgamma^-$ denote
the complimentary set.
We define three different types of moves, namely, add a covariate,
remove a covariate, and exchange a pair of covariates in and out of the
current model.
Each move starts by setting $\bgamma'=\bgamma$. Then we randomly
choose among the following:
\begin{itemize}
\item Add covariate: Generate $r \sim Q_{p-k}$,
and find the covariate $i \in\bgamma^-$ that has rank $r$ (among
covariates in $\bgamma^-$). Set $\bgamma'_i=1$.
\item Remove covariate uniformly: Uniformly pick $i \in\bgamma^+$,
and set $\bgamma'_i=0$.
\item Add a covariate and remove another: Pick $i$ uniformly from
$\bgamma^+$ and $j$ uniformly from $\bgamma^-$, and set $\bgamma
'_i=0; \bgamma'_j=1$.
\end{itemize}
In our current implementation, at each update we randomly select among
these moves with probabilities $0.45$, $0.45,$ and $0.1$.

\section{Calculations for Rao--Blackwellized estimates}\label{appmB}

In this appendix we derive the calculations need to compute the terms
in equation (\ref{post:inclusion}).

Let $\theta_{-j}$ denote the parameters $(\gamma_{-j},\bfb_{-j},
\tau, h, \pi)$. Note that
%
\begin{equation}
\Pr(\gamma_j = 1 | \bfy, \theta_{-j}) = \frac{\lambda}{1+\lambda},
\end{equation}
where
%
\begin{eqnarray} \label{eqn:postodds}
 \quad \lambda& :=& \frac{p(\gamma_j = 1 | \bfy, \theta_{-j})}{p(\gamma_j
= 0 | \bfy, \theta_{-j})} \nonumber\\
 \quad &\hspace*{3pt}=& \frac{p(\bfy| \gamma_j=1, \theta_{-j})} {p(\bfy| \gamma_j=0,
\theta_{-j})}
\frac{p(\bfb_{-j} | \gamma_j=1,\gamma_{-j},\tau,h,\pi)}{ p(\bfb
_{-j} | \gamma_j=0, \gamma_{-j},\tau,h,\pi)}
\frac{p(\gamma_j =1 | \gamma_{-j},\tau,h,\pi)} {p(\gamma_j =0
|\gamma_{-j},\tau,h,\pi)} \\
 \quad &\hspace*{3pt}=& \frac{p(\bfy| \gamma_j=1, \theta_{-j})} {p(\bfy| \gamma_j=0,
\theta_{-j})}
\frac{p(\bfb_{-j} | \gamma_j=1,\gamma_{-j},\tau,h)}{ p(\bfb_{-j}
| \gamma_j=0,\gamma_{-j},\tau,h)} \frac{\pi}{1-\pi}.\nonumber
\end{eqnarray}
The second term here arises because in our parameterization $\bfb
_{-j}$ is not independent of $\gamma_j$
(because its prior variance, $\sigma_a$, is a function of $h,\bgamma
$). This term is easily computed from the
fact that $\bfb_{-j} | \bgamma, \tau, h$ are i.i.d. $\sim N(0, \sigma
^2(h, \bgamma)/\tau)$.

To compute the numerator of the first term note that
%
\begin{equation}
\bfy|\gamma_j=1, \qquad  \theta_{-j} \sim N( X_{\bgamma-j}\beta_{\bgamma
-j} + \mu+ X_j \beta_j, 1/\tau I),
\end{equation}
with the priors on $\mu, \beta_j$ [from (\ref{prior:mutau})] being
%
\begin{eqnarray}
\mu| \tau&\sim& N(0, \sigma_{\mu}^2/\tau),\nonumber
\\[-8pt]
\\[-8pt]
\beta_j | \tau&\sim& N(0, \sigma_a^2/\tau).
\nonumber
\end{eqnarray}
Integrating out $\mu, \beta_j$ gives
%
\begin{equation} \qquad
\ \ \ p(\bfy|\gamma_j=1, \tau) = (2\pi)^{-n/2} \tau^{n/2} \frac{|\Omega
|^{1/2}}{\sigma_{\mu} \sigma_a} \exp
 \biggl( -\frac{1}{2} (R^t R - R^t X \Omega X^t R) \tau \biggr),
\end{equation}
where $R = \bfy- X_{\bgamma-j}\beta_{\bgamma-j}$, $\bgamma-j$
denotes the vector obtained by taking $\bgamma$ and setting the $j$th
coordinate to 0, $\Omega= (X^t X + \nu^{-1})^{-1}$, $\nu=
\left({\sigma_{\mu}^2 \atop 0} \enskip {0 \atop\sigma_a^2}
\right)
$, and $X = (1, X_j)$ is an $n \times2$ design matrix whose first
column is all 1s. [See equation (8) from Protocol S1 in \citet
{servinstephens07}.] The posterior distribution on $\beta_j$ is
given by
%
\begin{equation}
\beta_j | \bfy, \qquad  \theta_{-j} \sim N(\Omega X^t R, \Omega).
\end{equation}

Similarly, to compute the denominator of the first term, we use
%
\begin{equation}
\bfy|\gamma_j=0,  \qquad \theta_{-j} \sim N\bigl( X_{\bgamma-j}\beta_{\bgamma
-j} + \mu, (1/\tau) I\bigr),
\end{equation}
with priors on $\mu| \tau\sim N(0, \sigma_{\mu}^2/\tau)$.
Integrate out $\mu$ to get
%
\begin{equation}
\ \ \ p(\bfy|\gamma_j=0, \tau) = (2\pi)^{-n/2} \tau^{n/2} \frac{\Omega
_0^{1/2}}{\sigma_{\mu}} \exp
 \biggl ( -\frac{1}{2} (R^t R - \Omega_0 n^2 \bar{R}^2) \tau \biggr),
\end{equation}
where $\Omega_0 = (\sigma_{\mu}^{-2}+n)^{-1}$ and $\bar{R} = \frac
{1}{n}\sum{R_i}$.

From this we obtain
%
\begin{equation}
\frac{p(\bfy| \gamma_j=1, \theta_{-j})}{p(\bfy| \gamma_j=0,
\theta_{-j})}
= \frac{|\Omega|^{1/2}}{\Omega_0^{1/2}} \frac{1}{\sigma_a} \exp
  \biggl(\frac{\tau}{2}(R^t X\Omega X^tR-\Omega_0 n^2\bar
{R}^2) \biggr).
\end{equation}
In the limit $\sigma_{\mu} \rightarrow\infty$ we have $\Omega_0
\rightarrow n$ and $\nu\rightarrow
\left({0 \atop 0} \enskip {0 \atop\sigma_a^2} \right)
$ and the above expression becomes
%
\begin{equation}
\frac{p(\bfy| \gamma_j=1, \theta_{-j})}{p(\bfy| \gamma_j=0,
\theta_{-j})} =  |\Omega|^{1/2} \frac{n^{1/2}}{\sigma_a} \exp
 \biggl (\frac{\tau}{2}(R^t X\Omega X^tR-n\bar{R}^2) \biggr) .
\end{equation}

Note that this calculation effectively involves a univariate regression
of the residuals $R$ against covariate $j$. Furthermore, all covariates
$j \notin\bgamma^+$ use the same residuals: only for $j \in\bgamma
^+$ do the residuals need to be recomputed.
\end{appendix}

\section*{Acronyms used in the paper}
\begin{itemize}
\item BMA: Bayesian model averaging
\item BVSR: Bayesian variable selection regression
\item GWAS: genome wide association studies
\item LASSO: least absolute shrinkage and selection operator, a popular
variable selection method
\item MCMC: Markov chain Monte Carlo
\item PIP: posterior inclusion probability
\item PVE: proportion of variance explained
\item RPG: relative prediction gain
\item SNP: single nucleotide polymorphism
\item SIS: sure independence screen, a two-stage variable selection procedure.
\end{itemize}

\section*{Acknowledgments}

We thank two anonymous referees, and the editor and associate editor
for helpful comments on the initial submission.
We thank P. Carbonetto for useful discussions.



\printaddresses

\end{document}